\documentclass[acmsmall]{acmart}

\AtBeginDocument{%
  }

\usepackage[utf8]{inputenc} 
\usepackage[T1]{fontenc}    
\usepackage{url}            
\usepackage{booktabs}       
\usepackage{amsfonts}       
\usepackage{nicefrac}       
\usepackage{microtype}      
\usepackage{xcolor}         
\usepackage{graphicx}
\usepackage{amsmath}
\usepackage{tabularx}
\usepackage[linesnumbered, ruled]{algorithm2e}
\usepackage{multirow}
\usepackage{vcell}
\usepackage{listings}
\usepackage{enumitem}
\usepackage{makecell}
\usepackage{subcaption}
\usepackage{url}
\usepackage{siunitx}

\usepackage{colortbl}
\usepackage{lipsum}
\usepackage{tcolorbox}
\usepackage{arydshln}
\usepackage{lineno}
\usepackage{makecell}
\usepackage{bm}

\begin{document}


\title{IRCoCo: Immediate Rewards-Guided Deep Reinforcement Learning for Code Completion}

\author{Bolun Li}
\orcid{0009-0003-5006-6737}
\affiliation{%
  \institution{School of Information Science and Engineering, Shandong Normal University}
  \city{Jinan}
  \country{China}
}
\email{libolun118@gmail.com}

\author{Zhihong Sun}
\orcid{0009-0007-1387-3010}
\affiliation{%
  \institution{School of Information Science and Engineering, Shandong Normal University}
  \city{Jinan}
  \country{China}
}
\email{2022021002@stu.sdnu.edu.cn}

\author{Tao Huang}
\orcid{0009-0009-6955-7417}
\affiliation{%
  \institution{School of Information Science and Engineering, Shandong Normal University}
  \city{Jinan}
  \country{China}
}
\email{2022317095@stu.sdnu.edu.cn}

\author{Hongyu Zhang}
\orcid{0000-0002-3063-9425}
\affiliation{%
  \institution{Chongqing University}
  \city{Chongqing}
  \country{China}
}
\email{hyzhang@cqu.edu.cn}

\author{Yao Wan}
\orcid{0000-0001-6937-4180}
\affiliation{%
  \institution{Huazhong University of Science and Technology}
  \city{Wuhan}
  \country{China}
}
\email{wanyao@hust.edu.cn}

\author{Ge Li}
\orcid{0000-0002-5828-0186}
\affiliation{%
  \institution{Key Laboratory of High Confidence Software Technologies (Peking University), Ministry of Education; School of Computer Science, Peking University, Beijing}
  \country{China}
}
\email{lige@pku.edu.cn}

\author{Zhi Jin}
\orcid{0000-0003-1087-226X}
\authornote{Zhi Jin and Chen Lyu are the corresponding authors.}
\affiliation{%
  \institution{Key Laboratory of High Confidence Software Technologies (Peking University), Ministry of Education; School of Computer Science, Peking University, Beijing}
  \country{China}
}
\email{zhijin@pku.edu.cn}

\author{Chen Lyu}
\orcid{0000-0002-5044-1459}
\authornotemark[1]
\authornote{This work was done when Chen Lyu was a visiting scholar at Peking University.}
\affiliation{%
  \institution{School of Information Science and Engineering, Shandong Normal University}
  \city{Jinan}
  \country{China}
}
\email{lvchen@sdnu.edu.cn}


\begin{abstract}

  Code completion aims to enhance programming productivity by predicting potential code based on the current programming context. Recently, pre-trained language models (LMs) have become prominent in this field. Various approaches have been proposed to fine-tune LMs using supervised fine-tuning (SFT) techniques for code completion. However, the inherent \textit{exposure bias} of these models can cause errors to accumulate early in the sequence completion, leading to even more errors in subsequent completions. To address this problem, deep reinforcement learning (DRL) is an alternative technique for fine-tuning LMs for code completion, which can improve the generalization capabilities and overall performance. Nevertheless, integrating DRL-based strategies into code completion faces two major challenges: 1) The dynamic nature of the code context requires the completion model to quickly adapt to changes, which poses difficulties for conventional DRL strategies that focus on delayed rewarding of the final code state. 2) It is difficult to evaluate the correctness of partial code, thus the reward redistribution-based strategies cannot be adapted to code completion. To tackle these challenges, we propose IRCoCo, a code completion-specific DRL-based fine-tuning framework. This framework is designed to provide immediate rewards as feedback for detecting dynamic context changes arising from continuous edits during code completion. With the aid of immediate feedback, the fine-tuned LM can gain a more precise understanding of the current context, thereby enabling effective adjustment of the LM and optimizing code completion in a more refined manner. Experimental results demonstrate that fine-tuning pre-trained LMs with IRCoCo leads to significant improvements in the code completion task, outperforming both {SFT-based and other DRL-based baselines}. 
\end{abstract}


\setcopyright{acmlicensed}
\acmDOI{10.1145/3643735}
\acmYear{2024}
\copyrightyear{2024}
\acmSubmissionID{fse24main-p114-p}
\acmJournal{PACMSE}
\acmVolume{1}
\acmNumber{FSE}
\acmArticle{9}
\acmMonth{7}
\received{2023-09-29}
\received[accepted]{2024-01-23}


%




\maketitle

\section{Introduction}

Intelligent code completion can significantly boost the productivity of software developers by offering automated suggestions of subsequent code elements, based on the programming contexts (e.g., the already typed partial code). It has evolved into a key feature within contemporary integrated development environments (IDEs), exemplified by Visual Studio Code and its Copilot extension, and IntelliJ IDEA with its IntelliSense feature. Based on the degree of code completion, we classify contemporary code completion tools into two distinct categories:
1) token-level code completion, which centers on predicting individual tokens like function names, variable names, keywords, and operators within the code context~\cite{kim2021code, li2018code, wang2021code, svyatkovskiy2019pythia}; and 2) line-level code completion, designed to tackle the completion of multiple tokens, including function or class definitions, or the completion of multi-word expressions within intricate statements~\cite{wang2020towards,lu2022reacc,izadi2022codefill}. 
In this paper, we narrow our research scope to the latter one, which presents a more demanding challenge.

Recently, we have witnessed remarkable achievements in code generation, exemplified by the outstanding performance of pre-trained language models (LMs) as demonstrated by tools like GitHub Copilot~\cite{friedman2021introducing} and Amazon's CodeWhisperer~\cite{amazon2023ai}. However, pre-training an LM on a vast code corpus remains a time-consuming and computationally demanding endeavor, rendering it impractical for academic and small business environments with limited computing resources~\cite{huang2023not}.
Furthermore, subscribing to a code completion service may raise serious concerns about privacy leakage for many organizations. 
For instance, recent reports have revealed three incidents of data leakage at Samsung Electronics~\cite{report} when using online code completion services, e.g., ChatGPT.
These concerns underscore a pressing need to develop a localized and customized code completion model based on fine-tuning techniques for personal use.

To tackle the aforementioned concerns, many approaches based on supervised fine-tuning (SFT) have been proposed to refine LMs to the specific task of code completion, thereby enhancing their effectiveness within authentic code completion contexts~\cite{izadi2022codefill, liu2020multi}.
In the SFT of a code completion model, we typically refine the parameters of models by maximizing the log-likelihood of the subsequent ground-truth code token, also referred to as ``teacher-forcing''. We argue that the ``teacher-forcing'' strategy may suffer the \textit{exposure bias} issue. This issue emerges because, during the training phase, models are consistently exposed to ground-truth sequences. However, in the testing phase without ground truth, these models must predict based on their prior outputs, causing potential discrepancies between training and testing (i.e., \textit{exposure bias}). Over time, this \textit{exposure bias} would lead to accumulating errors in the testing phase, hindering the model from generating tokens that might be contextually appropriate but had a lower likelihood of being chosen during the training process~\cite{bengio2015scheduled, ranzato2016sequence}.

To mitigate the issue of \textit{exposure bias} inherent in ``teacher-forcing'', deep reinforcement learning (DRL), which utilizes a reward function to guide the model towards optimal completion sequences during training, is developed. DRL, rather than sequentially predicting tokens, blends exploration and exploitation. Through its dual-network design, with the ``actor'' offering localized token predictions and the ``critic'' giving global feedback on potential state outcomes, DRL refines decision-making. This combined strategy ensures that the model can recognize and utilize contextually appropriate tokens, even those with lower probabilities that might be overlooked by using the actor network in isolation. Thus,  the issue of \textit{exposure bias} can be resolved. On some related code intelligence tasks, such as natural language to code (NL2Code), DRL-based models show more promising results. For instance, Shojaee et al. \cite{shojaee2023execution} proposed the PPOCoder, which is a fine-tuned model, guided by the reward signals derived from unit tests conducted as code generation is completed. 
We refer to the reward obtained upon completion of code generation as a delayed reward. 
Le et al. \cite{le2022coderl} proposed CodeRL, which obtains a delayed reward based on whether the generated code can pass the unit tests. This delayed reward is subsequently redistributed to individual-generated tokens, reflecting their significance in achieving positive unit test outcomes.

Motivated by the aforementioned insights, this paper introduces an innovative DRL-based alignment method specifically designed for code completion. 
To maximize the potential of the DRL reward mechanism and guide the code completion model toward precise predictions, two pivotal challenges arise: 1) handling of dynamic intents and 2) evaluation of partial code. 

\noindent{\textbf{\textup{Challenge 1: Handling of Dynamic Intents.}}}
The intents for the code completion task are determined by the context of the code in the file currently being edited. As edits are made during the completion process, the context code changes, giving these intents a dynamic nature. It is difficult for existing delayed reward-based strategies, such as PPOCoder, to provide accurate feedback on such dynamically changing intents. Such strategies predominantly rely on the evaluative outcomes of the final generated code as a form of rewarding feedback to the environment. As a result, such an approach fails to capture the nuanced alterations in the context presented by intermediate code fragments during the completion process. These nuanced changes can perturb the code completion model's precise semantic understanding of the current code context and exert a substantial influence on the model's ensuing decisions. 

\noindent{\textbf{\textup{Challenge 2: Evaluation of Partial Code.}}}
Code completion typically produces partial code based on the local context. Since such code does not always offer complete functionality, it is challenging to directly perform unit testing to verify its correctness. Thus, reward redistribution-based strategies, such as those adopted by CodeRL~\cite{le2022coderl}, which rely on the analysis of unit test results of the generated code, cannot be adapted to code completion. Moreover, benchmark datasets designated for code completion (e.g., Py150~\cite{raychev2016probabilistic} and Java Corpus~\cite{allamanis2013mining}) often lack test cases, which further exacerbates the challenge of evaluating partial code.

These challenges lead to a substantial gap between the capabilities the code completion model aspires to achieve and the support contemporary DRL methodologies offer. 
To mitigate this gap, we propose IRCoCo (\textbf{I}mmediate \textbf{R}ewards-Guided Deep Reinforcement Learning for \textbf{Co}de \textbf{Co}mpletion), a DRL-based alignment mechanism that is model-agnostic, devised specifically for the unique demands of code completion tasks. \textbf{First}, to tackle \textit{Challenge 1}, we formulate an immediate rewards-guided DRL alignment architecture. This architecture offers real-time environmental feedback corresponding to the changing context during the code completion process. Such immediate feedback enables the model to more finely discern shifts in both intents and code semantics. Consequently, the model can iteratively adjust its generative strategy, thereby outputting code snippets that are in alignment with the current, most up-to-date intents. \textbf{Second}, acknowledging that code completion tasks often lack overt functional feedback mechanisms, such as unit tests, we design a novel reward shaping method based on the beneficial evaluation of subsequent code fragment completion to mitigate \textit{Challenge 2}. 
The core motivation behind this approach is to evaluate the validity of the tokens generated at a given time step for the accurate completion of subsequent code fragments. 
Within this dynamically evolving context, the immediate reward accurately gauges the extent to which the current token influences the ultimate code completion outcome. This immediate feedback facilitates more effective strategy learning for the model, harmonizes the balance between exploration and exploitation, and further refines the precision of subsequent token generation. This, in turn, accommodates the fluid nature of code completion contexts and enhances the accuracy of the code completion.

In this paper, we develop IRCoCo utilizing the actor-critic framework and 
examine the performance across six widely adopted pre-trained LMs.
These LMs serve as the actor to perform code completion, while two quality evaluators for generated code, trained using BLEU ~\cite{papineni2002bleu} and Edit-Sim  ~\cite{svyatkovskiy2020intellicode} metrics, act as the critic to produce immediate rewards; these components are incorporated within the IRCoCo framework. We assess IRCoCo's performance using two comprehensive datasets for Python and Java, culminating in 28 distinct experimental configurations (7 pre-trained LMs $\times$ 2 quality evaluators $\times$ 2 programming languages). The experimental outcomes indicate that across various configurations, IRCoCo consistently enhances the efficacy of code completion models. 
For instance, when employing CodeGPT as the actor and utilizing BLEU to train the code completion quality evaluator as the critic, the Edit-Sim scores demonstrate an improvement of \SI{7.9}{\percent} and \SI{1.6}{\percent} for Python and Java datasets, respectively, compared to the SFT method. Similarly, the EM scores exhibit significant enhancements of \SI{40.2}{\percent} and \SI{4.2}{\percent}, while the BLEU-4 scores see increases of \SI{14.7}{\percent} and \SI{1.6}{\percent}, compared to the SFT method. Additionally, the CodeBLEU~\cite{ren2020codebleu} score witness improvements of \SI{7.9}{\percent} and \SI{0.6}{\percent} in the same datasets, compared to the SFT method.

In summary, this paper makes the following major contributions.
\setlist[itemize]{left=0pt}
\begin{itemize}
    \item \textbf{Significant Problem.} We provide a comprehensive analysis of the characteristics of code completion tasks and identify effective ways to apply SFT and DRL-based alignment mechanisms to code completion.
    Our study introduces a novel, targeted fine-tuning paradigm specifically tailored for code completion tasks.
    \item \textbf{Novel Approach.} We introduce IRCoCo, an immediate rewards-guided DRL framework with great potential to improve the performance of pre-trained LMs on code completion.
    \item \textbf{Extensive Experiments.} 
    We perform extensive
    experiments using six open-source pre-trained LMs on the Py150 and Java Corpus datasets. Our empirical findings show that our methodology yields models 
    with significantly improved performance across various metrics, including Edit-Sim, EM, CodeBLEU, and BLEU,
    thereby substantially improving code completion capabilities. 
\end{itemize}

\section{MOTIVATION}
\subsection{A Motivating Example}
\begin{figure*}[!htb]
    \centering

    \begin{subfigure}[t]{0.48\linewidth}
        \includegraphics[width=\linewidth]{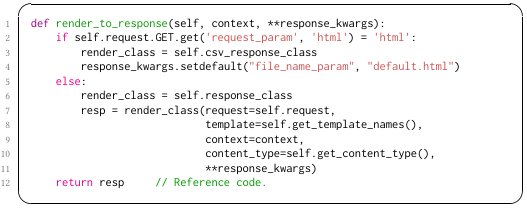}
        \caption{The source code.}
        \label{fig:top}
    \end{subfigure}
    \hfill
    \begin{minipage}[t]{0.48\linewidth}
        \vspace{-76pt} 

        \begin{subfigure}[t]{\linewidth}
            \includegraphics[width=\linewidth]{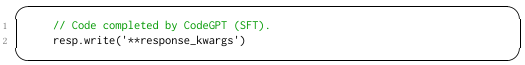}
            \caption{Code completed by CodeGPT (SFT).}
            \label{fig:left}
        \end{subfigure}
        
        \vspace{6pt} 

        \begin{subfigure}[b]{\linewidth}
            \includegraphics[width=\linewidth]{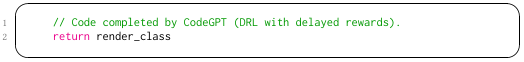}
            \caption{Code completed by CodeGPT (DRL w/ delayed rewards).}
            \label{fig:right}
        \end{subfigure}
    \end{minipage}
    \caption{The developer-written code, completed by CodeGPT trained by SFT, completed by CodeGPT trained by DRL w/ delayed rewards.}
    \label{fig:combined}
    \vspace{-1em}
\end{figure*}

In Figure~\ref{fig:combined}, we use an example to illustrate the motivation of our work.
Figure~\ref{fig:top} shows a function \texttt{render\_to\_response} that is randomly sampled from the Py150 dataset~\cite{raychev2016probabilistic}.
This function operates by selecting the appropriate rendering class according to request parameter values, utilizing that class to generate a response object, and ultimately returning the object.
Figures~\ref{fig:left} and~\ref{fig:right} present the code completion results generated by CodeGPT using two distinct training strategies: SFT and DRL with delayed reward. These results are generated based on an incomplete code fragment, specifically the code spanning lines 1 to 11 as depicted in Figure~\ref{fig:top}.
From these examples, we can derive the following observations.

As shown in Figure~\ref{fig:left}, it is clear to see that the code completed by CodeGPT based on SFT strategy deviates significantly from the reference code. 
In SFT, the model is fine-tuned using the ``teacher-forcing'' strategy.
However, this approach introduces significant discrepancies between the training and inference phases, resulting in the emergence of the \textit{exposure bias} issue during inference. 
To illustrate, when the anticipated output is \texttt{return}, the model may produce the \texttt{resp} object instead. Consequently, this initial error will propagate to subsequent completions, influencing the incorrect invocation of the \texttt{.write()} function, even its relevance to the current context is negligible.




In terms of the DRL-based strategy with delayed rewards, as shown in Figure~\ref{fig:right}, CodeGPT, when trained using this approach, occasionally exhibits errors.
While we can see a performance advancement compared to that achieved using the SFT strategy, and the model seemingly comprehends the intended action following the \texttt{return}, it still lacks complete accuracy. More specifically, in the reference code, the \texttt{render\_class} object is assigned to a variable named \texttt{resp}; nevertheless, the model fails to accurately capture this modification.
The error could be ascribed to the limitations inherent in traditional DRL methods, which provide feedback to the model only after it completes the entire sequence. 
As a result, the model fails to allocate rewards during intermediate states, possibly leading to erroneous decisions, especially when feedback is delayed until the end of the generated sequence. Furthermore, the repetitive occurrence of the \texttt{render\_class} in the subsequent code may divert the model's attention, causing it to overlook more optimal solutions.
\subsection{Key Ideas}
From the example above, we identify two essential features a proficient code completion model needs to possess, serving as the inspiration for our proposed approach: 1) the ability to mitigate error accumulation during code completion; 2) the capability to perceive the latest context changes.
The key idea of our approach centers around the development of a DRL framework guided by immediate rewards, based on the actor-critic framework~\cite{bahdanau2016actor}. 
In this framework, we assign the role of the actor network to the code completion model, typically an LM, which undergoes fine-tuning through DRL, with the goal of alleviating the inherent \textit{exposure bias} in SFT, thereby mitigating error accumulation during code completion.
Concurrently, we develop and train a critic network to assess the quality of code completion. This critic network provides immediate rewards for each code token generated by the actor network, 
effectively converting the sparse feedback in DRL, resulting from delayed rewards, into dense feedback. This ensures the timely integration of the latest contextual information, thereby minimizing delays in learning. 
The code completion model is further refined during the policy gradient optimization phase inherent in DRL. 
Next, we will detail the fundamentals of this designed strategy.

The interaction between actor and critic networks can be analogized to the relationship between a student and a teacher. In this analogy, the actor assumes the role of the student, while the critic serves as the teacher.
From the actor's perspective, the reward signals from the critic play a pivotal role in enhancing its ability to discern the ever-evolving context in code completion, enabling the actor to make prompt adjustments based on immediate feedback. 
Specifically, the actor feeds each completed token to the critic, which subsequently assigns a score to assess the quality of the token in the current context. This scoring mechanism provides valuable insights to the actor, aiding in the refinement of its token completion strategy.

From the critic's perspective, the task is to assign a reasonable reward to every token produced by the actor.
Nevertheless, crafting effective immediate rewards for each generated code token poses a significant challenge.
Previously, CodeRL~\cite{le2022coderl} derived reward signals based on the code's ability to pass unit tests. 
Subsequently, the model quantifies the impact of each generated token on the overall outcome of the code's unit test results and reallocates the delayed reward across individual tokens proportionally to their respective contributions.
However, employing this approach in code completion is impractical for two main reasons: 1) The constraint arises from the frequent absence of test cases in datasets for code completion, rendering code completions incapable of deriving reward signals from unit tests. 
2) There is currently no established and effective method for assessing the contribution of a specific portion of code within the dynamic context of code completion.
The primary insight regarding the critic is that by furnishing a justifiable reward for each token generated by the actor, the actor can gain a heightened sensitivity to the dynamically evolving context throughout its training.
This empowers the actor to assess the effectiveness of the generated code segment in shaping the final outcome. Adopting such an approach facilitates a more refined balance between exploration and exploitation, thereby enhancing the actor's performance.

\subsection{Feasibility Analysis}
A foundational premise of our approach is that the immediate reward given by the critic to each token generated by the actor serves as feedback, indicating if the token aids in the subsequent code completion.
Acquiring these rewards poses a considerable challenge; however, in contrast to code generation, code completion involves shorter sequences and is not contingent upon the code's execution outcome. This distinction substantially alleviates the complexity and cost associated with calculating these immediate rewards.
Considering the inherent features of code completion, we propose to utilize the widely adopted evaluation metrics (e.g., BLEU and Edit-Sim) to design methods for assessing immediate rewards. 
Specifically, we suggest assigning higher rewards to tokens if the completion results generated by the actor, based on the current token, exhibit closer alignment with the correct outcomes. This encourages the actor to promptly adjust its strategy, promoting the generation of tokens with an increased likelihood of yielding correct completions.

In addition, we examine the influence of this score on guiding the model towards generating accurate subsequent completion sequences. 
Experimental results demonstrate that our method surpasses 
those DRL methods with delayed rewards.
On the Py150 dataset, our proposed immediate rewards demonstrate notable improvements across various metrics. Specifically, we observe average enhancements of \SI{3.32}{\percent}, \SI{4.56}{\percent}, \SI{3.75}{\percent}, and \SI{2.81}{\percent} in Edit-Sim, EM, BLEU-4, and CodeBLEU, respectively (see Table~\ref{tab:table2}). These results affirm the efficacy of the designed immediate rewards, showcasing their simplicity and effectiveness in significantly enhancing the performance of pre-trained LMs in code completion tasks.

\section{IRCoCo}
\label{gen_inst}
\begin{figure*}[htbp]
    \centering
    \includegraphics[width=0.8\textwidth]{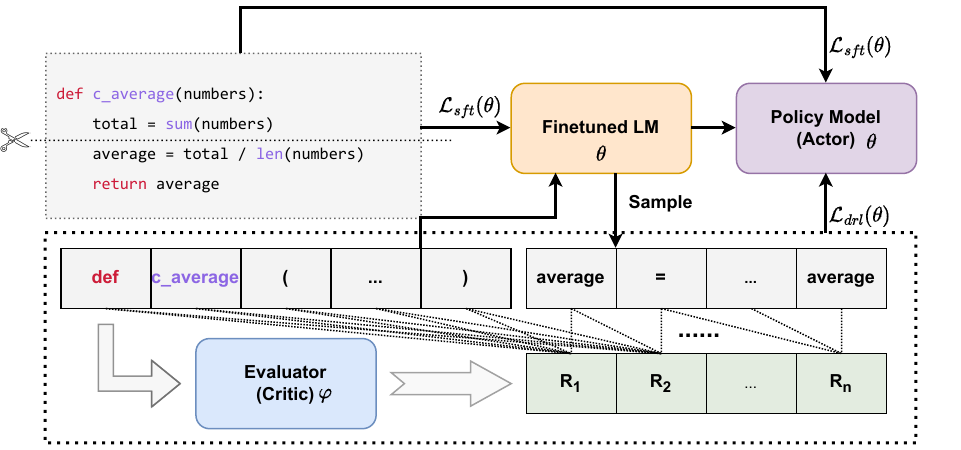}
    \caption{\textbf{Overview of the IRCoCo using the Actor-Critic Framework.} First, the actor network samples synthetic samples. These samples are generated token by token and are sequentially added to the end of the incomplete code fragment. Afterward, they are rewarded by the critic. Leveraging these immediate rewards, the strategy is refined by integrating the IRCoCo framework, which employs a joint fine-tuning approach using SFT and DRL.
    }
    \label{RL}
    \vspace{-1em}
\end{figure*}

The IRCoCo framework utilizes a combination of SFT and DRL to fine-tune the code completion model, aiming to improve accuracy and ensure both syntactic and semantic correctness of the generated code.
Figure~\ref{RL} shows an overview of the IRCoCo framework. The first step involves utilizing SFT method to fine-tune the pre-trained LM, which serves as the actor network for sampling synthetic samples. Subsequently, an evaluator model is trained as the critic network, responsible for evaluating the synthetic samples and returning reward scores. Finally, the LM (actor) is jointly optimized using SFT and DRL. In the following subsections, we will provide a detailed description of each component in the IRCoCo framework.

\subsection{Code Completion Task}
The goal of the code completion task studied in this paper is to predict the subsequent code fragments given a partial code context until the end-of-sequence special token \texttt{</s>} is generated. 
Specifically, given a partial code sequence $X=\{x_1,x_2,\ldots,x_k\}$, the task is to predict the following code fragments $Y=\{y_1,y_2,\ldots,y_n\}$ using an LM $p$ for code completion. Formally, the code completion task can be formulated as:
\begin{equation}
    p\left(Y \mid X\right)=\prod_{t=1}^{n}{p\left(y_t\mid y_{1:t-1},X\right)\ } \,.
\end{equation}

\subsection{Supervised Fine-Tuning-Based Model Training}
Typically, code completion is modeled as a sequence-to-sequence task
whose goal is to map the input sequence $X$ to the output sequence $Y=\{y_1,y_2,\ldots,y_n\}$, where each token $y_i$ is sampled from the vocabulary ($\mathcal{V}$) of code.
During the training period, the code completion model aims to minimize the cross entropy between the generated code and the reference code, based on the following training loss:
\begin{equation}
\label{equ:2}
	\mathcal{L}_{sft}(\theta)=-\sum_{t} \log p_\theta(Y \mid X)=-\sum_{t} \log \left[p_\theta\left(y_{t} \mid y_{1:t-1}, X\right)\right]\,,
\end{equation}
where $y_t$ is the output of each decoding step $t$, and $\theta$ is the model parameter.

\subsection{Quality Evaluator for Generated Code}
In the IRCoCo framework, the evaluator assesses the quality of code completions given an incomplete code fragment, enabling code completion models to meticulously sense dynamically changing contextual requirements. Drawing inspiration from \cite{sun2023don}, we adopt the same Transformer-based GPT-2 model as the primary architecture for the evaluator, due to its pre-training on a large-scale corpus and demonstrated success in NL generation tasks. To enhance the efficiency and performance of the evaluator, we restrict the model parameters to 16 million and incorporate a linear head layer into the Transformer-based GPT-2 architecture. To evaluate the similarity between the generated code fragments and target code, we employ BLEU and Edit-Sim respectively as the optimization metric for the evaluator.

\begin{figure*}[!t]
	\centering
	\includegraphics[width=0.72\textwidth]{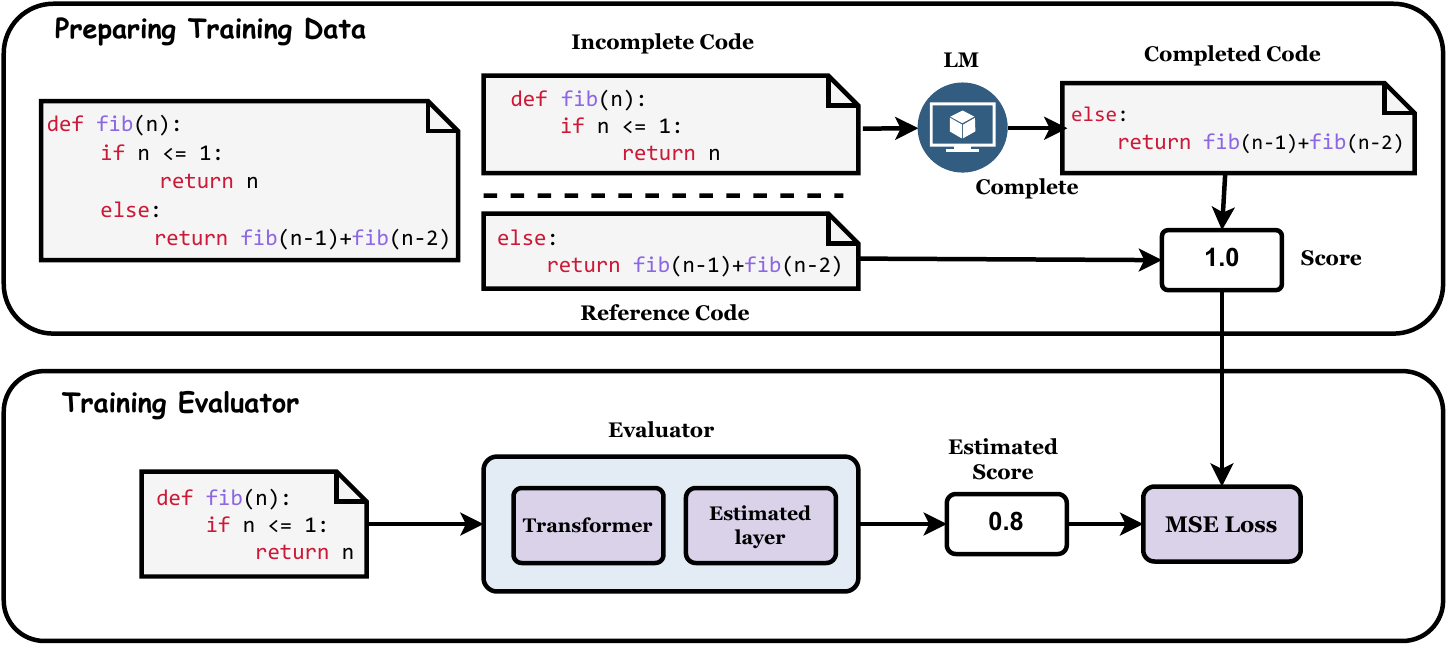}
 \vspace{-2mm}
	\caption{\textbf{Overview of the Evaluator.} Training the evaluator first requires preparing training data. In the training data preparation phase, we randomly split the complete code to obtain the incomplete code and reference code fragments. After that, we pass the incomplete code fragment through the LM to obtain the completed code and compute the score $s$. Finally, we pair the incomplete code fragment with the score $s$ to obtain the training data. In the training phase, we will obtain the score $s'$ by the evaluator, and the training goal is to minimize the MSE loss of $s$ and $s'$.
 }
	\label{evaluate}
     \vspace{-1em}
\end{figure*}

Our basic idea is to align the reward score distribution predicted by the evaluator with the discrepancy between the LM's completions and the reference code, as measured by the BLEU or Edit-Sim metrics. With such a training approach, the evaluator is capable of scoring any fragment of code. This score directly reveals the likelihood that, starting from the last token of the incomplete code fragment, the completion matches the reference code. A higher score indicates a greater expectation that the current token will lead to a correct completion. Unlike SFT, the optimization objective for each immediate reward focuses on the anticipated benefit from the entire completion, not just the local gain from the next token. With the guidance of this reward mechanism, the LM's prediction capabilities undergo continual refinement, aiming to bolster the likelihood of generating precise code.
Figure~\ref{evaluate} illustrates the training framework of the quality evaluator using BLEU as an optimization metric.

\noindentparagraph{\textbf{\textup{Preparing the Training Data.}}}
To train the quality evaluator, the first step involves obtaining training data. The effectiveness and performance of the evaluator are contingent on the quality and diversity of the training data. Thus, it is imperative to obtain representative code fragments from a large-scale codebase to use as training data. To obtain the necessary training data, we randomly divide a given complete code fragment $C$ into two parts. The first portion is treated as an incomplete code fragment $C_x$, while the second part serves as the corresponding reference code fragment $C_y$. Next, the incomplete code fragment $C_x$ is inputted into the fine-tuned LM, which generates the completed code fragment $C_g$. The generated code fragment $C_g$ is then compared to the reference code fragment $C_y$, and the accuracy is calculated to obtain the score $s$. By following this process, we can obtain the datasets for training the evaluator of code completion quality after pairing each incomplete code fragment $C_x$ with its corresponding score $s$.

\noindentparagraph{\textbf{\textup{Training the Evaluator.}}}
We utilize the standard GPT-2 architecture for our model. To process the training data ($C_x$, $s$), we employ a multi-headed attention mechanism in each Transformer block to enhance the linguistic representation by aggregating the previous output. Subsequently, we transform the output of the multiple attention heads into the final score $s^\prime $ using a linear layer:
\begin{equation}
	h_0=H_xW_{embedding}+W_{position}\,,
\end{equation}
\begin{equation}
    h_l = \text{TransformerBlock}\left(h_{l-1}\right),l\in [1,L]\,,
\end{equation}
\begin{equation}
    s^\prime=\text{Linear}\left(h_l\right)\,,
\end{equation}
where $H_x$ is the context vector of the incomplete code fragment $C_x$; $W_{embedding}$ is the token embedding matrix, $W_{position}$ is the position embedding matrix, and $h_0$ is the hidden state vector representation of the incomplete code fragment $ C_x$; $L$ is the number of layers of the transformer block, and $h_l$ is the hidden state vector representation of the incomplete code fragment $C_x$ in the $L$-th layer of the model; $Linear$ is the linear layer. $s^\prime$ is the score of the evaluator output.

The code completion quality evaluator minimizes the mean-square error between $s^\prime$ and $s$ as the final training goal:
\begin{equation}
\label{equ:6}
    \text{MSE}\left ( s,s^\prime \right ) =\frac{1}{N} \sum_{i=1}^{N}\left ( s_i^\prime-s_i \right )^{2}\,.
\end{equation}

\subsection{Reinforcement Learning-Based Alignment of Generated Code}
\label{sec:3.4}
To tackle the \textit{exposure  bias} issue, several studies~\cite{shojaee2023execution,dou2024stepcoder} have attempted to utilize DRL to train code completion models. Nevertheless, current DRL methods are prone to the delayed reward problem. The optimal policy may require multiple steps before attaining the maximum reward, leading to difficulties for the model to determine the optimal policy and a tendency to converge to local optimal solutions. For this reason, we propose to incorporate immediate rewards into DRL. Specifically, the code completion generation process is viewed as a Markov Decision Process (MDP)~\cite{bellman1957markovian} consisting of four main components:

\noindentparagraph{\textbf{\textup{State.}}} 
During each time step $t$ of the decoding process, the state $s_t$ consists of the incomplete code fragment $X$ and the word $\hat{y}_{1:t-1}$ generated earlier in the decoding process, i.e., $s_t=\left \{X,\hat{y}_{1:t-1}  \right \}$. In the initial state of decoding, the state $s_t$ consists of only the incomplete code fragment $X$, i.e., $\left \{X\right \}$. We use the hidden state vector $h_t$ as the vector representation under state $s_t$. 

\noindentparagraph{\textbf{\textup{Action.}}} 
In our scenario, the task involves predicting the subsequent code token by sampling a token ($\hat{y}_t$) from the vocabulary ($\mathcal{V}$) at each time step.
We conceptualize the task of predicting the next token as the action of sampling a token from a predefined vocabulary (action space).


\noindentparagraph{\textbf{\textup{Reward.}}} 
Rewards are used to evaluate whether the completed code facilitates the generation of subsequent completions. In this study, we aim to incentivize the code completion model to produce tokens that facilitate subsequent completions. To achieve this, we provide rewards to the model based on an evaluator (i.e., critic) that has been trained for this purpose. Therefore, we define the reward for each time step $t$ as:
\begin{equation}\label{equ:7}
r\left(s_{t}, \hat{y}_{t}\right)=\left\{
\begin{aligned}
& Q_\varphi \left(X;\hat{y}_{1:t-1} , \hat{y}_t\right) &  & \textbf{if}\ \hat{y}_t\neq \texttt{</s>} &\\
& Q_\varphi \left(X;\hat{y}_{1:t-1}\right) &  & \textbf{if}\  \hat{y}_t=\texttt{</s>} &
\end{aligned}
\right. 
\end{equation}
where $Q_\varphi$ is the trained evaluator. In particular, in this work, we give a valid reward for each generated time step $t$ of the code completion model and give the same reward for the final end-of-sequence special token \texttt{</s>} as for the previous token of the generation.

\begin{algorithm}[!t]
\small
\SetKwData{Left}{left}\SetKwData{This}{this}\SetKwData{Up}{up} \SetKwFunction{Union}{Union}\SetKwFunction{FindCompress}{FindCompress} \SetKwInOut{Input}{Input}\SetKwInOut{Output}{Output}
	\caption{The training process of IRCoCo}
    \label{alg:my_algorithm}
	\Input{A set of incomplete code and reference code pairs $(X, Y)$, along with the pre-trained LM $p$.} 
	\Output{The parameters of the model after fine-tuning $\theta$.}
	 \BlankLine 
	 
        Fine-tuning of actor $p_\theta$ using Eq. (\ref{equ:2})\; 
         Sampling synthetic samples $\hat{Y}$ using Eq. (\ref{equ:8})\; 
        Training critic $Q_\varphi$ using Eq. (\ref{equ:6})\; 
	 \For{number of epochs until convergence}{ 
            \For{$\left ( x,y \right ) \subset \left ( X,Y \right )$ and $\left ( x,\hat{y} \right ) \subset  ( X,\hat{Y} )$ }{
                \Repeat{ number of samples}{
                    \text{ \tcp{Calculate Reward}}\\
                    \text{Compute ${r} \left(s_t,\hat{y}_t\right)$ using Eq. (\ref{equ:7})}\\
                    \text{\tcp{Calculate Loss}}\\
                    $\mathcal{L}_{sft}\left(\theta\right)\gets -\sum_{t} \log p_\theta(Y \mid X)=-\sum_{t} \log \left[p_\theta\left(y_{t} \mid y_{1:t-1}, X\right)\right]$\\
                    $\mathcal{L}_{drl}\left(\theta\right)\gets -\mathbb{E}_{\hat{Y}\sim p_\theta} \left [  r(X, \hat{Y}) \right ]$\\
                    $\mathcal{L}\left(\theta\right){\gets\mathcal{L}_{sft}\left(\theta\right)+\mathcal{L}}_{drl}\left(\theta\right)$\\
                    \text{\tcp{Updata Model Parameters}}\\
                    $\theta\gets\theta-{\nabla}_\theta\mathcal{L}\left(\theta\right)$\\
                }
            }
    }       
\end{algorithm}
\noindentparagraph{\textbf{\textup{Policy.}}} 
The policy function $p_\theta\left(\hat{y}_t\middle| s_t\right)$ takes the current state $s_t$ as input and outputs $\hat{y}_t$ as the probability of the next completion token. In this work, we use the policy gradient~\cite{bahdanau2016actor} method to optimize the policy function. The definition of the policy function $p_\theta\left(\hat{y}_t\middle| s_t\right)$ is as follows:
\begin{equation}
\label{equ:8}
p_\theta\left(\hat{y}_t\middle| s_t\right)=p_\theta\left(\hat{y}_t\middle| \hat{y}_{1:t-1},X\right)=softmax\left(Kh_t+b\right)\,,
\end{equation}
where $K$ is the weight parameter of the model, $b$ is the bias vector, and $h_t$ is the hidden state at time step $t$.

The parameters of the code completion model $\theta$ can be thought of as stochastic strategies and the goal of model training is to find a policy network $p_\theta(\hat{Y}| X)$ to minimize negative expected returns:
\begin{equation}
\mathcal{L}_{drl}\left(\theta\right)=-\mathbb{E}_{\hat{Y}\sim p_\theta} \left [  r(X, \hat{Y}) \right ]\,,
\end{equation}
where $\hat{Y}=\left(\hat{y}_1,\ldots,\hat{y}_t\right)$ 
represents a sequence of synthetic samples, with each code token $\hat{y}_t$ being sampled by the code completion model at decoding time step $t$.
According to the DRL algorithm and policy gradient definition, we define the gradient $\mathbf{\nabla}_\theta \mathcal{L}_{drl}\left(\theta\right)$ of the non-differentiable regression reward function $r$ as:
\begin{equation}\label{1}
    \begin{aligned}
    \mathbf{\nabla}_\theta\mathcal{L}_{drl}\left(\theta\right)&\approx-\mathbb{E}_{\hat{Y}\sim p_\theta} \left [  r(X, \hat{Y})\mathbf{\nabla}_\theta\log p_\theta(\hat{Y} \mid X) \right ]
    \\
    &\approx-\mathbb{E}_{\hat{Y}\sim p_\theta}\left [ \sum_{t}r(X, \hat{Y})\mathbf{\nabla}_\theta\log p_\theta\left ( \hat{y}_t|\hat{y}_{1:t-1},X  \right ) \right ]  \,.
    \end{aligned}
\end{equation}
During the training process, we adopt a hybrid learning method based on SFT-based fine-tuning and DRL-based alignment, setting this as our final training objective: 
\begin{equation}
\mathcal{L}\left(\theta\right){=\mathcal{L}_{sft}\left(\theta\right)+\mathcal{L}}_{drl}\left(\theta\right)\,.
\end{equation}
SFT and DRL-based alignment have distinct advantages, with SFT enabling supervised learning through large-scale data and DRL allowing autonomous learning through intelligent trial and error. By integrating these two learning methods, we can leverage their strengths to enhance the model's performance and generalization. The ultimate loss function is determined by summing the losses of SFT-based and DRL-based alignment. 

Algorithm~\ref{alg:my_algorithm} presents the pseudo-code of training IRCoCo. For reinforcement learning data, we use the policy network $p$ to sample one synthetic sample $(X, \hat{Y})$ for each source code pair $(X, Y)$. Subsequently, we compute immediate reward based on the reward function defined in Section~\ref{sec:3.4} to estimate the payoff. Finally, we update the gradients based on the corresponding loss functions.

\section{Experimental Setup}
We have conducted several experiments to evaluate IRCoCo. Specifically, we seek to answer the following Research Questions (RQs).
\begin{itemize}
\item 
\textbf{RQ1: Effectiveness of Code Completion.} To what extent can the training process of IRCoCo help improve the capabilities of code completion models? To answer this question, we compare the performance of pre-trained LMs before and after integration with IRCoCo.

\item 
\textbf{RQ2: Validity of Immediate Rewards.} Are immediate rewards in the IRCoCo framework effective? To answer this question, we compare it with delayed reward-based DRL and several rule-based reward construction methods.

\item 
\textbf{RQ3: Impact of Different Model Learning Objectives.} Are the learning objectives for joint training in IRCoCo effective? To answer this question, we compare IRCoCo with SFT-only and DRL-only training modes.

\item 
\textbf{RQ4: Quantitative Analysis.} How does IRCoCo effective across varying token counts? To answer this question, we evaluate completions across different numbers of tokens.

\item 
\textbf{RQ5: Qualitative Analysis.} How realistic is the predictive power of IRCoCo? To answer this question, 
we conduct a qualitative analysis study on IRCoCo.
\end{itemize}

\subsection{Evaluation Datasets}

\begin{table}[!t]
\captionsetup{skip=0pt}
	\caption{Statistical analysis of the two datasets.}
	\label{tab:table1}
	\resizebox{\columnwidth}{!}{
    	\begin{tabular}{lcccc}
    		\toprule
    		Dataset & Examples & Average tokens of inputs
    		& Average token of outputs & Average lines of code \\
    		\midrule
      
    		Py150 & 50,000 & 96.9 & 9.06 & 11.6  \\
    		Java Corpus & 8,268 & 111.9 & 10.4 & 8.2 \\
    		\bottomrule
    	\end{tabular}
    }
    \label{table2}
    \vspace{-1em}
\end{table}
In our experiments, we adopt the Py150 dataset and the Java Corpus dataset, both widely used to evaluate code completion tasks.
The Py150 dataset~\cite{raychev2016probabilistic} comprises 150,000 code files in Python~2, which are partitioned into a training set of 100,000 files and a test set of 50,000 files. The Java Corpus dataset~\cite{allamanis2013mining} comprises almost 30,000 Java files, which are divided into a training set of 12,934 files and a test set of 8,268 files. 
Following the data preprocessing approach in the CodeXGLUE benchmark~\cite{lu2021codexglue}, we normalize the 200 most commonly used strings and the 30 most commonly used numeric characters by special tokens such as \texttt{<STR LIT:utf-8>} and \texttt{<NUM LIT>}.
As the original Py150 and Java Corpus datasets consist of complete code snippets, and given our objective to perform line-level code completion, 
we randomly cut the code data to reflect the diversity in real-world application scenarios. 
In our work, we randomly divide a code fragment into two parts: the first half served as an incomplete code fragment, and the second half comprised 10 tokens that are designated as reference code. 
During DRL-based alignment, we organize the data in a format identical to that of the APPS program synthesis benchmark~\cite{hendrycks2021measuring}, whereby each incomplete code is paired with one reference code and one sampled example. Our aim is to train the model to complete the next 10 tokens. The statistics of the data are presented in Table~\ref{table2}.

\subsection{Baselines} 
We utilize the following pre-trained LMs, which are widely used in code completion \cite{kim2021code,lu2022reacc,lu2021codexglue,li2023starcoder,wang2023codet5plus}, as the underlying base models to evaluate their performance both prior to and following the integration of the IRCoCo framework.

\setlist[itemize]{left=0pt}
\begin{itemize}
\item \textbf{GPT-2 [\SI{124}{M} \& \SI{1.5}{B}]:} GPT-2~\cite{radford2019language} is a pre-trained LM that harnesses the Transformer architecture. It undergoes pre-training via large-scale unsupervised learning and demonstrates robust performance on generative tasks, such as question answering and code completion.

\item \textbf{CodeGPT [\SI{124}{M}]:} CodeGPT~\cite{lu2021codexglue} is a Transformer-based code completion model with the same architecture and training objectives as GPT-2. It comprises a 12-layer Transformer decoder and is pre-trained using the Python and Java corpus of the CodeSearchNet~\cite{husain2019codesearchnet} datasets. Through this pre-training, the model acquires a comprehensive understanding of code structure and syntax rules, enabling it to automatically generate code.

\item \textbf{CodeGPT-adapt [\SI{124}{M}]:} CodeGPT-adapt~\cite{lu2021codexglue} is a domain adaptive model, which is trained on the code corpus with the same vocabulary as GPT-2 as a starting point, and inherits the natural language understanding capability of GPT-2.

\item \textbf{CodeGen [\SI{350}{M}]:} CodeGen~\cite{nijkamp2022codegen} uses a standard Transformer-based autoregressive LM with a next-token prediction LM as a learning objective, trained on NL and PL datasets, and has demonstrated excellent performance in the field of program synthesis.

\item \textbf{StarCoder [\SI{164}{M}]:} StarCoder~\cite{li2023starcoder} is based on the GPT architecture, which is obtained after training based on the licensed data on GitHub, and we use StarCoder [164M] in our experiments, which has the same architecture as StarCoder. 

\item \textbf{CodeT5+ [\SI{220}{M}]:} 
CodeT5+~\cite{wang2023codet5plus} is an encoder-decoder-based masked language model, utilizing diverse training tasks and a simple yet effective pre-training method in its pre-training phase, offering enhanced support for program comprehension and code completion compared to CodeT5.
\end{itemize}

\subsection{Evaluation Metrics} 
Following previous studies \cite{wang2020towards,lu2022reacc,lu2021codexglue}, we employ Edit Similarity (Edit-Sim)~\cite{svyatkovskiy2020intellicode}, BLEU-4 \cite{papineni2002bleu} similarity metric, CodeBLEU~\cite{ren2020codebleu} metric, and Exact Match Accuracy (EM)~\cite{svyatkovskiy2020intellicode} to evaluate IRCoCo.

\subsection{Implementation Details} 
For the six baselines considered in our work, we follow existing works~\cite{lu2021codexglue,wang2023codet5plus,wang2021codet5} that experiment on CodeXGLUE and fine-tune the pre-trained LM in Huggingface \cite{wolf2019huggingface}. Subsequently, these fine-tuned models undergo the DRL-based alignment process.
The experiments involving GPT-2 [\SI{1.5}{B}] are conducted using an Nvidia GeForce RTX A6000 GPU with \SI{48}{GB} memory. Other experiments are performed on a server with two Nvidia GeForce RTX 3090 GPUs with \SI{22}{GB} memory.

The process of SFT is elaborated as follows. For the decoder-only model, both the input and output consist of complete code fragments. In contrast, for the encoder-decoder model, the input comprises incomplete code fragments, while the output represents the corresponding code to be completed. Specifically, at each time step $t$, a ``teacher-forcing'' strategy is used and the next correct code token is generated based on the first $t-1$ tokens in the reference code.

For the code completion quality evaluator (i.e., critic network), we train two evaluators using the BLEU and Edit-Sim indicators respectively to provide rewards for the code completion model. Specifically, we use the Transformer-based GPT-2 model, setting the number of layers to 4, the number of heads to 4, the embedding size of code token to 256, and the epochs to 30.

In terms of DRL, we update the parameters of the code completion model (i.e., actor network) once for each batch. The hyperparameters employed during DRL-based alignment are consistent with those used in the SFT-based process. 
Specifically, we have defined the experimental parameters as follows: a batch size of 2, a learning rate set at $2 \times e^{-5}$, and a total of 10 epochs.



\section{Experimental Results}

\subsection{Effectiveness of Code Completion (RQ1)} 
Table~\ref{tab:table10} presents the performance of IRCoCo with various base pre-trained LMs on two evaluation datasets. It is evident that IRCoCo significantly enhances code completion performance, regardless of whether the evaluator is trained with BLEU or Edit-Sim metrics. For instance, in the Py150 dataset, there is an increase of approximately \SI{7}{\percent} in Edit-Sim scores, around \SI{40}{\percent} in EM scores, around \SI{6}{\percent} in CodeBLEU scores, and close to \SI{10}{\percent} in BLEU-4 scores. More specifically, when CodeGPT undergoes training using the code completion quality evaluator guided by the BLEU metric, its Edit-Sim score rises from \SI{60.66}{\percent} to \SI{65.44}{\percent}, marking an increment of \SI{7.9}{\percent}. Concurrently, its EM score ascends from \SI{15.65}{\percent} to \SI{21.94}{\percent}, a notable surge of \SI{40.2}{\percent}, while the BLEU-4 score elevates from \SI{38.10}{\percent} to \SI{43.71}{\percent}, a rise of \SI{14.7}{\percent}. Additionally, the CodeBLEU metric shows a significant improvement, climbing from \SI{42.42}{\percent} to \SI{45.79}{\percent}, an increase of \SI{7.9}{\percent}. In the Java Corpus dataset, we observe enhancements of about \SI{2}{\percent} in Edit-Sim, roughly \SI{8}{\percent} in EM scores, and nearly \SI{2}{\percent} in BLEU-4 scores. Furthermore, there is an approximate increase of \SI{2.3}{\percent} in the CodeBLEU scores. It is worth noting that metrics recorded on Java dataset are generally inferior to those on Python dataset. This disparity can be attributed to the lengthier nature of Java code; on average, Java code snippets comprise 112 tokens, in contrast to Python's 97 tokens.

\begin{table*}
    \setlength{\abovecaptionskip}{0pt}%
    \setlength{\belowcaptionskip}{10pt}%
    \renewcommand{\arraystretch}{1.2}
    \setlength{\tabcolsep}{2pt} 
	\caption{Performance of IRCoCo with various base pre-trained LMs on two evaluation datasets (in \%). }
	\label{tab:table10}
	\resizebox{1.0\columnwidth}{!}{
	    \begin{tabular}{lccccccccccccccccc}
		    \toprule
		    \multirow{3}{*}{\textbf{Model}} & \multicolumn{8}{c}{BLEU (Evaluator) }  & \multicolumn{8}{c}{Edit-Sim (Evaluator)}\\  
      \cmidrule(r){2-9} \cmidrule(r){10-17}  & \multicolumn{4}{c}{Py150} & \multicolumn{4}{c}{Java Corpus} & \multicolumn{4}{c}{Py150} & \multicolumn{4}{c}{Java Corpus}  \\
		    \cmidrule(r){2-5} \cmidrule(r){6-9} \cmidrule(r){10-13} \cmidrule(r){14-17}
		    & Edit-Sim & EM & BLEU-4 & CodeBLEU  & Edit-Sim & EM & BLEU-4 & CodeBLEU &Edit-Sim & EM & BLEU-4 & CodeBLEU  & Edit-Sim & EM & BLEU-4 & CodeBLEU   \\
		    \midrule
		    GPT-2 [124M] & 59.37 & 9.21 & 35.31 & 40.81 & 57.42 & 4.95  & 32.56 & 42.25 & 59.37 & 9.21 & 35.31 & 40.81 &  57.42 & 4.95  & 32.56 & 42.25 \\
		    \rowcolor{gray!=20}GPT-2 [124M]+IRCoCo & \textbf{63.65} & \textbf{13.93}  & \textbf{39.96} & \textbf{43.70}   & \textbf{58.58} & \textbf{5.40} & \textbf{33.12} & \textbf{43.68} &  \textbf{63.87} & \textbf{14.12} & \textbf{40.87} & \textbf{43.83} & \textbf{58.37} & \textbf{5.33} & \textbf{33.06} & \textbf{43.86}  \\
      
            Relative Improvement & (\textcolor{red}{$\uparrow \SI{7.2}{\percent}$}) & (\textcolor{red}{$\uparrow \SI{51.2}{\percent}$})  & (\textcolor{red}{$\uparrow \SI{13.2}{\percent}$})&
            (\textcolor{red}{$\uparrow \SI{7.1}{\percent}$})& (\textcolor{red}{$\uparrow \SI{2.0}{\percent}$})& (\textcolor{red}{$\uparrow \SI{9.1}{\percent}$})& (\textcolor{red}{$\uparrow \SI{1.7}{\percent}$})& 
            (\textcolor{red}{$\uparrow \SI{3.4}{\percent}$})& (\textcolor{red}{$\uparrow \SI{7.6}{\percent}$})& (\textcolor{red}{$\uparrow \SI{53.3}{\percent}$})& (\textcolor{red}{$\uparrow \SI{15.7}{\percent}$})& 
            (\textcolor{red}{$\uparrow \SI{6.2}{\percent}$})& (\textcolor{red}{$\uparrow \SI{1.7}{\percent}$})& (\textcolor{red}{$\uparrow \SI{7.7}{\percent}$})& (\textcolor{red}{$\uparrow \SI{1.5}{\percent}$})&
            (\textcolor{red}{$\uparrow \SI{3.8}{\percent}$}) \\
            \hdashline

            GPT-2 [1.5B] & 65.62 & 14.45 & 41.93 & 44.76 & 58.14 & 6.24  & 34.24 & 43.47 & 65.62 & 14.45 & 41.93 & 44.76 & 58.14 & 6.24  & 34.24 & 43.47 \\

            \rowcolor{gray!=20}GPT-2 [1.5B]+IRCoCo & \textbf{66.90} & \textbf{16.37}  & \textbf{43.32} & \textbf{45.91}   & \textbf{59.51} & \textbf{6.69} & \textbf{35.02} & \textbf{43.98} &  \textbf{66.50} & \textbf{16.26} & \textbf{43.25} & \textbf{45.81} & \textbf{59.26} & \textbf{6.53} & \textbf{34.87} & \textbf{43.76}  \\

            Relative Improvement & (\textcolor{red}{$\uparrow \SI{2.0}{\percent}$}) & (\textcolor{red}{$\uparrow \SI{13.3}{\percent}$})  & (\textcolor{red}{$\uparrow \SI{3.3}{\percent}$})&
            (\textcolor{red}{$\uparrow \SI{2.6}{\percent}$})& (\textcolor{red}{$\uparrow \SI{2.4}{\percent}$})& (\textcolor{red}{$\uparrow \SI{7.2}{\percent}$})& (\textcolor{red}{$\uparrow \SI{2.3}{\percent}$})& 
            (\textcolor{red}{$\uparrow \SI{1.2}{\percent}$})& (\textcolor{red}{$\uparrow \SI{1.3}{\percent}$})& (\textcolor{red}{$\uparrow \SI{12.5}{\percent}$})& (\textcolor{red}{$\uparrow \SI{3.1}{\percent}$})& 
            (\textcolor{red}{$\uparrow \SI{2.3}{\percent}$})& (\textcolor{red}{$\uparrow \SI{1.9}{\percent}$})& (\textcolor{red}{$\uparrow \SI{4.6}{\percent}$})& (\textcolor{red}{$\uparrow \SI{1.8}{\percent}$})&
            (\textcolor{red}{$\uparrow \SI{0.6}{\percent}$}) \\
            \hdashline

            CodeGPT & 60.66 & 15.65 & 38.10 & 42.42 & 58.98 & 12.06\ & 36.32 & 43.92 & 60.66 & 15.65 & 38.10 & 42.42 & 58.98 & 12.06\ & 36.32 & 43.92 \\
             \rowcolor{ gray!20}CodeGPT+IRCoCo &\textbf{65.44} & \textbf{21.94} & \textbf{43.71} &\textbf{45.79} & \textbf{59.94} & \textbf{12.57} & \textbf{36.89} &\textbf{44.19} &   \textbf{65.93} & \textbf{22.32} & \textbf{43.96} &\textbf{45.53} & \textbf{59.91} & \textbf{12.77} & \textbf{36.94} &\textbf{44.32}  \\

             Relative Improvement & (\textcolor{red}{$\uparrow \SI{7.9}{\percent}$}) & (\textcolor{red}{$\uparrow \SI{40.2}{\percent}$})  & (\textcolor{red}{$\uparrow \SI{14.7}{\percent}$})& 
             (\textcolor{red}{$\uparrow \SI{7.9}{\percent}$})&(\textcolor{red}{$\uparrow \SI{1.6}{\percent}$})& (\textcolor{red}{$\uparrow \SI{4.2}{\percent}$})& (\textcolor{red}{$\uparrow \SI{1.6}{\percent}$})& 
             (\textcolor{red}{$\uparrow \SI{0.6}{\percent}$})&(\textcolor{red}{$\uparrow \SI{8.7}{\percent}$})& (\textcolor{red}{$\uparrow \SI{42.6}{\percent}$})& (\textcolor{red}{$\uparrow \SI{15.4}{\percent}$})&
             (\textcolor{red}{$\uparrow \SI{7.3}{\percent}$})&
             (\textcolor{red}{$\uparrow \SI{1.6}{\percent}$})& (\textcolor{red}{$\uparrow \SI{5.9}{\percent}$})& (\textcolor{red}{$\uparrow \SI{1.7}{\percent}$})&
             (\textcolor{red}{$\uparrow \SI{0.9}{\percent}$})\\
             \hdashline
             
		   CodeGPT-adapt & 63.08  & 13.10 & 39.31 & 43.27 & 58.87 & 5.57 & 33.53 & 42.97 & 63.08  & 13.10 & 39.31 & 43.27 & 58.87 & 5.57 & 33.53 & 42.97 \\
	       \rowcolor{ gray!20}CodeGPT-adapt+IRCoCo &  \textbf{63.68} & \textbf{14.02} & \textbf{40.15} & \textbf{43.63} & \textbf{59.15} & \textbf{5.84} & \textbf{33.68} & \textbf{44.02} &  \textbf{64.05} & \textbf{14.32} & \textbf{40.86} & \textbf{43.96} & \textbf{59.32} & \textbf{6.05} & \textbf{33.98} & \textbf{43.75}  \\

            Relative Improvement & (\textcolor{red}{$\uparrow \SI{1.0}{\percent}$}) & (\textcolor{red}{$\uparrow \SI{7.0}{\percent}$})  & (\textcolor{red}{$\uparrow \SI{2.1}{\percent}$}) & (\textcolor{red}{$\uparrow \SI{0.8}{\percent}$}) & (\textcolor{red}{$\uparrow \SI{0.5}{\percent}$})& (\textcolor{red}{$\uparrow \SI{4.8}{\percent}$})& (\textcolor{red}{$\uparrow \SI{0.4}{\percent}$}) & (\textcolor{red}{$\uparrow \SI{2.4}{\percent}$}) & (\textcolor{red}{$\uparrow \SI{1.5}{\percent}$})& (\textcolor{red}{$\uparrow \SI{9.3}{\percent}$})& (\textcolor{red}{$\uparrow \SI{4.0}{\percent}$}) & (\textcolor{red}{$\uparrow \SI{1.6}{\percent}$}) & (\textcolor{red}{$\uparrow \SI{0.8}{\percent}$})& (\textcolor{red}{$\uparrow \SI{8.6}{\percent}$})& (\textcolor{red}{$\uparrow \SI{1.3}{\percent}$}) & (\textcolor{red}{$\uparrow \SI{1.8}{\percent}$})  \\
            \hdashline
            
        CodeGen & 59.45  & 10.18 & 35.53 & 40.90 & 59.33 & 11.52 & 35.96 & 43.08 & 59.45  & 10.18 & 35.53 & 40.90 & 59.33 & 11.52 & 35.96 & 43.08  \\
	       \rowcolor{ gray!20}CodeGen+IRCoCo &  \textbf{64.55} & \textbf{14.46} & \textbf{41.31} &  \textbf{43.08} & \textbf{60.31} & \textbf{12.40} & \textbf{36.99} &  \textbf{44.06} &  \textbf{64.03} & \textbf{14.21} & \textbf{40.97} &  \textbf{42.86} & \textbf{60.17} & \textbf{12.22} & \textbf{36.87} &  \textbf{43.81} \\

            Relative Improvement & (\textcolor{red}{$\uparrow \SI{8.6}{\percent}$}) & (\textcolor{red}{$\uparrow \SI{42.0}{\percent}$})  & (\textcolor{red}{$\uparrow \SI{16.3}{\percent}$}) & (\textcolor{red}{$\uparrow \SI{5.3}{\percent}$}) & (\textcolor{red}{$\uparrow \SI{1.7}{\percent}$})& (\textcolor{red}{$\uparrow \SI{7.6}{\percent}$})& (\textcolor{red}{$\uparrow \SI{2.9}{\percent}$}) & (\textcolor{red}{$\uparrow \SI{2.3}{\percent}$}) & (\textcolor{red}{$\uparrow \SI{7.7}{\percent}$})& (\textcolor{red}{$\uparrow \SI{39.6}{\percent}$})& (\textcolor{red}{$\uparrow \SI{15.3}{\percent}$}) & (\textcolor{red}{$\uparrow \SI{4.8}{\percent}$}) & (\textcolor{red}{$\uparrow \SI{1.4}{\percent}$})& (\textcolor{red}{$\uparrow \SI{6.1}{\percent}$})& (\textcolor{red}{$\uparrow \SI{2.5}{\percent}$}) & (\textcolor{red}{$\uparrow \SI{1.7}{\percent}$})  \\
        \hdashline
        
        StarCoder & 61.37  & 16.44 & 39.11 & 43.11 & 59.94 & 11.88 & 36.24 & 43.56 & 61.37  & 16.44 & 39.11 & 43.11 & 59.94 & 11.88 & 36.24 & 43.56  \\
	       \rowcolor{ gray!20}StarCoder+IRCoCo &  \textbf{64.02} & \textbf{20.38} & \textbf{42.06} & \textbf{44.76} & \textbf{61.63} & \textbf{13.15} & \textbf{37.96} & \textbf{44.38} &  \textbf{64.37} & \textbf{20.73} & \textbf{42.32} & \textbf{45.49} & \textbf{61.08} & \textbf{12.96} & \textbf{37.41} & \textbf{44.71} \\

            Relative Improvement & (\textcolor{red}{$\uparrow \SI{4.3}{\percent}$}) & (\textcolor{red}{$\uparrow \SI{24.0}{\percent}$})  & (\textcolor{red}{$\uparrow \SI{7.5}{\percent}$}) & (\textcolor{red}{$\uparrow \SI{3.8}{\percent}$}) & (\textcolor{red}{$\uparrow \SI{2.8}{\percent}$})& (\textcolor{red}{$\uparrow \SI{10.7}{\percent}$})& (\textcolor{red}{$\uparrow \SI{4.8}{\percent}$}) & (\textcolor{red}{$\uparrow \SI{1.8}{\percent}$}) & (\textcolor{red}{$\uparrow \SI{4.9}{\percent}$})& (\textcolor{red}{$\uparrow \SI{26.1}{\percent}$})& (\textcolor{red}{$\uparrow \SI{8.2}{\percent}$}) & (\textcolor{red}{$\uparrow \SI{5.5}{\percent}$}) & (\textcolor{red}{$\uparrow \SI{1.9}{\percent}$})& (\textcolor{red}{$\uparrow \SI{9.1}{\percent}$})& (\textcolor{red}{$\uparrow \SI{3.3}{\percent}$})& (\textcolor{red}{$\uparrow \SI{2.6}{\percent}$})  \\
        \hdashline
        
        CodeT5+ & 55.81  & 5.33 & 32.99 & 38.71 & 53.42 & 4.61 & 31.49 & 36.85 & 55.81  & 5.33 & 32.99 & 38.71 & 53.42 & 4.61 & 31.49 & 36.85  \\
	       \rowcolor{ gray!20}CodeT5+ +IRCoCo&  \textbf{56.97} & \textbf{6.25} & \textbf{33.74} & \textbf{39.96} & \textbf{54.11} & \textbf{4.98} & \textbf{32.06} & \textbf{38.61} &  \textbf{57.33} & \textbf{6.46} & \textbf{33.98} & \textbf{39.21} & \textbf{54.58} & \textbf{5.13} & \textbf{32.34} & \textbf{39.86}  \\

            Relative Improvement & (\textcolor{red}{$\uparrow \SI{2.1}{\percent}$}) & (\textcolor{red}{$\uparrow \SI{17.3}{\percent}$})  & (\textcolor{red}{$\uparrow \SI{2.3}{\percent}$}) & (\textcolor{red}{$\uparrow \SI{3.2}{\percent}$}) & (\textcolor{red}{$\uparrow \SI{1.3}{\percent}$})& (\textcolor{red}{$\uparrow \SI{8.0}{\percent}$})& (\textcolor{red}{$\uparrow \SI{1.8}{\percent}$}) & (\textcolor{red}{$\uparrow \SI{4.8}{\percent}$}) & (\textcolor{red}{$\uparrow \SI{2.7}{\percent}$})& (\textcolor{red}{$\uparrow \SI{21.2}{\percent}$})& (\textcolor{red}{$\uparrow \SI{3.0}{\percent}$}) & (\textcolor{red}{$\uparrow \SI{1.3}{\percent}$}) & (\textcolor{red}{$\uparrow \SI{2.2}{\percent}$})& (\textcolor{red}{$\uparrow \SI{11.3}{\percent}$})& (\textcolor{red}{$\uparrow \SI{2.7}{\percent}$}) & (\textcolor{red}{$\uparrow \SI{8.2}{\percent}$})  \\
		 
		    \bottomrule
	\end{tabular}
}
\vspace{-1em}
\end{table*}  

One surprising finding is that CodeGPT-adapt without adding the IRCoCo framework outperforms CodeGPT. However, after adding the IRCoCo framework, the metrics of CodeGPT-adapt+IRCoCo are lower than that of CodeGPT+IRCoCo. This is attributed to the inherent properties of the pre-trained LM itself. CodeGPT-adapt is a domain adaptive model, and the primary objective of domain adaptive models is to transfer learning between different data distributions and are highly sensitive to changes in the data distribution. On the other hand, SFT and DRL-based alignment have different objectives, which may result in insignificant performance improvements for the model. Overall, incorporating the pre-trained LM into the IRCoCo framework generally improves its performance in code generation. This suggests that utilizing the immediate rewards provided by the IRCoCo framework enables the model to acquire better strategies.

\begin{tcolorbox}  
\textbf{Answer to RQ1:} Our results indicate that incorporating the pre-trained LM into the IRCoCo framework generally improves its performance in code generation. This suggests that utilizing the immediate rewards provided by the IRCoCo framework enables the model to acquire better strategies, leading to improved effectiveness.
\end{tcolorbox}

\subsection{Validity of Immediate Rewards (RQ2)} 
To evaluate the effectiveness of the immediate rewards, we investigate three different reward shaping strategies. For delayed rewards (DR), we allocate rewards exclusively at the end of code completion (i.e., after the generation of the end-of-sequence token \texttt{</s>}), assigning 0 to intermediate tokens. For immediate rewards, our first approach employs a linear attenuation (LA) based rule \cite{le2022coderl}, which decays rewards based on token position, ranging from time steps $t=1$ to $t=T$. The second strategy involves a binary (0-1) reward system, where we evaluate the completion status of each token. If the generated token is exactly aligned with the reference code token, a reward of 1 is awarded, otherwise a value of 0 is assigned. The experimental results are shown in Table~\ref{tab:table2}.

\begin{table*}[!t]
    \setlength{\abovecaptionskip}{0pt}%
    \setlength{\belowcaptionskip}{10pt}%
    \setlength{\tabcolsep}{2pt} 

    \renewcommand{\arraystretch}{1.1}
	\caption{Comparative results of our proposed method with delayed rewards (DR), linearly attenuating (LA) rewards, and 0-1 (0-1) based rewards.}
	\label{tab:table2}
	\resizebox{1.0\columnwidth}{!}{
	    \begin{tabular}{lccccccccccccccccc}
		    \toprule
		    \multirow{3}{*}{\textbf{Model}} & \multicolumn{8}{c}{BLEU (Evaluator)}  & \multicolumn{8}{c}{Edit-Sim (Evaluator)}\\  
      \cmidrule(r){2-9} \cmidrule(r){10-17}  & \multicolumn{4}{c}{Py150} & \multicolumn{4}{c}{Java Corpus} & \multicolumn{4}{c}{Py150} & \multicolumn{4}{c}{Java Corpus}  \\
		    \cmidrule(r){2-5} \cmidrule(r){6-9} \cmidrule(r){10-13} \cmidrule(r){14-17}
		    & Edit-Sim & EM & BLEU-4 & CodeBLEU  & Edit-Sim & EM & BLEU-4 & CodeBLEU &Edit-Sim & EM & BLEU-4 & CodeBLEU  & Edit-Sim & EM & BLEU-4 & CodeBLEU   \\
		    \midrule
		    GPT-2 [124M] (DR) & 57.05 & 9.05 & 35.55 & 40.57 & 56.92 & 5.27  & 32.98 & 43.02 & 57.65 & 9.43 & 35.81 & 40.68 & 57.82 & 5.21  & 32.90 & 42.74  \\
            GPT-2 [124M] (LA) & 62.57 & 13.15 & 39.22 & 43.62 & 57.59 & 5.29  & 32.30 & 43.11 & 62.57 & 13.15 & 39.22 & 43.62 & 57.59 & 5.29  & 32.30 & 43.11 \\
            GPT-2 [124M] (0-1) & 58.34 & 8.14 & 33.33 & 40.09 & 57.11 & 4.46  & 32.38 & 42.93 & 58.34 & 8.14 & 33.33 & 40.09 & 57.11 & 4.46  & 32.38 & 42.93 \\
		    \rowcolor{gray!=20}GPT-2 [124M]+IRCoCo & \textbf{63.65} & \textbf{13.93} & \textbf{39.96} & \textbf{43.70} & \textbf{58.58} & \textbf{5.40} & \textbf{33.12} & \textbf{43.68} &  \textbf{63.87} & \textbf{14.12} & \textbf{40.87} & \textbf{43.83} & \textbf{58.37} & \textbf{5.33} & \textbf{33.06} & \textbf{43.86}   \\
		    \hdashline

        GPT-2 [1.5B] (DR) & 65.11 & 14.26 & 41.88 & 44.64 & 57.62 & 5.97  & 33.96 & 43.18 & 65.02 & 14.13 & 41.79 & 44.42 & 57.47 & 5.91  & 33.84 & 43.06  \\

        GPT-2 [1.5B] (LA) & 65.68 & 15.37 & 42.08 & 45.18 & 58.46 & 6.37  & 34.44 & 43.62 & 65.68 & 15.37 & 42.08 & 45.18 & 58.46 & 6.37  & 34.44 & 43.62  \\

        GPT-2 [1.5B] (0-1) & 66.16 & 15.61 & 42.33 & 45.30 & 58.17 & 6.21  & 34.28 & 43.53 & 66.16 & 15.61 & 42.33 & 45.30 & 58.17 & 6.21  & 34.28 & 43.53 \\

        \rowcolor{gray!=20}GPT-2 [1.5B]+IRCoCo & \textbf{66.90} & \textbf{16.37} & \textbf{43.32} & \textbf{45.91} & \textbf{59.51} & \textbf{6.69} & \textbf{35.02} & \textbf{43.98} &  \textbf{66.50} & \textbf{16.26} & \textbf{43.25} & \textbf{45.81} & \textbf{59.26} & \textbf{6.53} & \textbf{34.87} & \textbf{43.76}   \\
        \hdashline

        CodeGPT (DR) & 62.12 & 17.38 & 39.92 & 42.98 & 52.07 & 7.42 & 34.29 & 42.16 & 62.30 & 17.52 & 40.08 & 43.14 & 53.26 & 8.05 & 34.93 & 42.89 \\
        CodeGPT (LA) & 64.45 & 21.03 & 42.84 & 45.03 & 59.18 & 11.72 & 36.01 & 44.02 & 64.45 & 21.03 & 42.84 & 45.03 & 59.18 & 11.72 & 36.01 & 44.02 \\
        CodeGPT (0-1) & 63.18 & 17.11 & 38.94 & 42.60 & 58.46 & 11.28  & 35.34 & 43.77  & 63.18 & 17.11 & 38.94 & 42.60 & 58.46 & 11.28  & 35.34 & 43.77 \\
        \rowcolor{ gray!20}CodeGPT+IRCoCo &\textbf{65.44} & \textbf{21.94} & \textbf{43.71} &\textbf{45.79} & \textbf{59.94} & \textbf{12.57} & \textbf{36.89} &\textbf{44.19} &  \textbf{65.93} & \textbf{22.32} & \textbf{43.96} &\textbf{45.53} & \textbf{59.91} & \textbf{12.77} & \textbf{36.94} &\textbf{44.32}  \\
        \hdashline
             
		CodeGPT-adapt (DR) & 62.78  & 13.14 & 39.28 & 43.09 & 58.40 & 5.53 & 33.32 & 43.37 & 63.46  & 13.48 & 39.60 & 43.33 & 58.24 & 5.41 & 33.28 & 43.12  \\
        CodeGPT-adapt (LA) & 63.34 & 13.60 & 39.55 & 42.94 & 58.50 & 5.26 & 32.94 & 43.37 & 63.34 & 13.60 & 39.55 & 42.94 & 58.50 & 5.26 & 32.94 & 43.37 \\
        CodeGPT-adapt (0-1) & 62.91 & 11.91 & 38.89 & 42.15 & 58.61 & 5.03 & 32.46 & 42.75 & 62.91 & 11.91 & 38.89 & 42.15 & 58.61 & 5.03 & 32.46 & 42.75 \\
	    \rowcolor{ gray!20}CodeGPT-adapt+IRCoCo &  \textbf{63.68} & \textbf{14.02} & \textbf{40.15} &  \textbf{43.63} & \textbf{59.15} & \textbf{5.84} & \textbf{33.68} &  \textbf{44.02} &  \textbf{64.05} & \textbf{14.32} & \textbf{40.86} &  \textbf{43.96} & \textbf{59.32} & \textbf{6.05} & \textbf{33.98} &  \textbf{43.75}  \\
        \hdashline

        CodeGen (DR) & 60.33  & 10.71 & 36.24 & 40.61 & 56.84 & 9.61 & 34.71 & 41.15 & 60.06  & 10.55 & 36.09 & 40.37 & 57.22 & 9.88 & 34.89 & 41.83  \\
        CodeGen (LA) & 62.46  & 12.89 & 38.41 & 41.32 & 58.92 & 11.10 & 35.34 & 42.66 & 62.46  & 12.89 & 38.41 & 41.32 & 58.92 & 11.10 & 35.34 & 42.66 \\
        CodeGen (0-1) & 61.44 & 11.26 & 37.18 & 41.43 & 59.16 & 11.86  & 36.03 & 42.74 & 61.44 & 11.26 & 37.18 & 41.43 & 59.16 & 11.86  & 36.03 & 42.74 \\
	    \rowcolor{ gray!20}CodeGen+IRCoCo &  \textbf{64.55} & \textbf{14.46} & \textbf{41.31} & \textbf{43.08} & \textbf{60.31} & \textbf{12.40} & \textbf{36.99} & \textbf{44.06} &  \textbf{64.03} & \textbf{14.21} & \textbf{40.97} & \textbf{42.86} & \textbf{60.17} & \textbf{12.22} & \textbf{36.87} & \textbf{43.81} \\
        \hdashline
        
        StarCoder (DR) & 61.51  & 16.56 & 39.23 & 42.78 & 57.56 & 10.08 & 35.17 & 42.63 & 61.86  & 16.85 & 39.55 & 42.91 & 56.44 & 9.53 & 34.61 & 42.01 \\
        StarCoder (LA) & 62.77  & 17.01 & 39.86 & 44.16 & 60.44 & 12.23 & 36.89 & 43.85 & 62.77  & 17.01 & 39.86 & 44.16 & 60.44 & 12.23 & 36.89 & 43.85 \\
        StarCoder (0-1) & 61.96 & 15.48 & 38.77 & 43.45 & 60.81 & 11.67  & 36.02 & 44.02 & 61.96 & 15.48 & 38.77 & 43.45 & 60.81 & 11.67  & 36.02 & 44.02 \\
	    \rowcolor{ gray!20}StarCoder+IRCoCo &  \textbf{64.02} & \textbf{20.38} & \textbf{42.06} &  \textbf{44.76} & \textbf{61.63} & \textbf{13.15} & \textbf{37.96} &  \textbf{44.38} &  \textbf{64.37} & \textbf{20.73} & \textbf{42.32} &  \textbf{45.49} & \textbf{61.08} & \textbf{12.96} & \textbf{37.41} &  \textbf{44.71}  \\
        \hdashline
        
        CodeT5+(DR) & 55.42  & 5.11 & 32.69 & 38.14 & 53.51 & 4.20 & 31.38 & 36.47 & 55.16  & 5.22 & 32.52 & 38.26 & 53.71 & 4.58 & 31.30 & 36.05  \\
        CodeT5+(LA) & 55.13  & 5.01 & 31.77 & 38.23 & 53.29 & 4.44 & 31.08 & 37.17 & 55.13  & 5.01 & 31.77 & 38.23 & 53.29 & 4.44 & 31.08 & 37.17  \\
        CodeT5+ (0-1) & 54.36 & 4.28 & 30.24 & 37.66 & 52.66 & 3.23  & 30.26 & 35.46 & 54.36 & 4.28 & 30.24 & 37.66 & 52.66 & 3.23  & 30.26 & 35.46 \\
	    \rowcolor{ gray!20}CodeT5+ +IRCoCo&  \textbf{56.97} & \textbf{6.25} & \textbf{33.74} & \textbf{39.96} & \textbf{54.11} & \textbf{4.98} & \textbf{32.06} & \textbf{38.61} &  \textbf{57.33} & \textbf{6.46} & \textbf{33.98} & \textbf{39.21} & \textbf{54.58} & \textbf{5.13} & \textbf{32.34} & \textbf{39.86}  \\

		    \bottomrule
	\end{tabular}
}
\vspace{-1em}
\end{table*} 

From the table, it is evident that IRCoCo consistently surpasses the three reward design methods across all metrics. For instance, in the Py150 dataset, when employing CodeGPT as the code completion model and a quality evaluator trained by BLEU, IRCoCo's performance exceeds that of DR by approximately \SI{3.5}{\percent} in Edit-Sim, \SI{4.5}{\percent} in EM, \SI{3}{\percent} in CodeBLEU, and \SI{4}{\percent} in BLEU-4. This indicates that, in contrast to the delayed reward-based approach, IRCoCo can adeptly discern dynamically shifting contextual demands. Furthermore, compared to the other two immediate reward configurations, IRCoCo exhibits enhancements across multiple metrics. This underscores the potential of rewards derived from our code completion quality evaluator in motivating the model to generate superior subsequent completion sequences.

\begin{tcolorbox}  
\textbf{Answer to RQ2:} Our results indicate that IRCoCo can adeptly discern dynamically shifting contextual demands. Furthermore, compared to the other two immediate reward configurations, IRCoCo exhibits enhancements across multiple metrics. This underscores the validity of immediate rewards.
\end{tcolorbox}
\begin{table}[!t]
\centering
\captionsetup{skip=0pt}
\caption{\textbf{Results with different learning objectives.} `$LM$' indicates that LM has completed 5 rounds of SFT. `$LM$ + $\mathcal{L}_{sft}$' is an additional 5 rounds of SFT based on `$LM$'; `$LM$ + $\mathcal{L}_{drl}$' is 10 rounds of DRL-based alignment based on '$LM$'; `$LM$ + $\mathcal{L}_{sft}$ + $\mathcal{L}_{drl}$' is a 10-round joint training of SFT and DRL-based alignment based on `$LM$'.}
\renewcommand{\arraystretch}{1.3}
\setlength{\abovecaptionskip}{0pt}%
\setlength{\belowcaptionskip}{0pt}%
\setlength{\tabcolsep}{2pt} 

\resizebox{\columnwidth}{!}{
\label{table:table3}
\begin{tabular}{l|cccc|cccc|cccc|cccc}
\hline
\multicolumn{17}{c}{Py150} \\
\hline
\multirow{2}{*}{Model} & \multicolumn{4}{c|}{$LM$} & \multicolumn{4}{c|}{$LM$ + $\mathcal{L}_{sft}$} & \multicolumn{4}{c|}{$LM$ + $\mathcal{L}_{drl}$} & \multicolumn{4}{c}{$LM$ + $\mathcal{L}_{sft}$ + $\mathcal{L}_{drl}$} \\
\cline{2-17}
 & Edit-Sim & EM & BLEU-4 & CodeBLEU & Edit-Sim & EM & BLEU-4 & CodeBLEU & Edit-Sim & EM & BLEU-4 & CodeBLEU & Edit-Sim & EM & BLEU-4 & CodeBLEU \\
\hline
GPT-2 [124M] & 59.37 & 9.21 & 35.31 & 40.81 & 59.48 & 9.00 & 35.37 & 40.69 & 58.15 & 8.86 & 34.98 & 40.26 & \textbf{63.65} & \textbf{13.93} & \textbf{39.96} & \textbf{43.70} \\
GPT-2 [1.5B] & 65.62 & 14.45 & 41.93 & 44.76 & 62.36 & 12.23 & 39.45 & 42 .82 & 63.74 & 12.51 & 39.87 & 44.06 & \textbf{66.90} & \textbf{16.37} & \textbf{43.32} & \textbf{45.91} \\
CodeGPT & 60.66 & 15.65 & 38.10 & 42.42 & 58.47 & 12.57 & 35.30 & 41.17 & 58.72 & 12.72 & 38.82 & 43.10 & \textbf{65.44} & \textbf{21.94} & \textbf{43.71} & \textbf{45.79} \\
CodeGPT-adapt & 63.08 & 13.10 & 39.31 & 43.27 & 60.84 & 10.75 & 36.67 & 41.73 & 61.51 & 10.96 & 37.30 & 42.34 & \textbf{63.68} & \textbf{14.02} & \textbf{40.15} & \textbf{43.63} \\
CodeGen & 59.45 & 10.18 & 35.53 & 40.90 & 58.33 & 9.52 & 34.88 & 40.22 & 58.82 & 9.81 & 35.16 & 41.43 & \textbf{64.55} & \textbf{14.46} & \textbf{41.31} & \textbf{43.08}  \\
StarCoder & 61.37 & 16.44 & 39.11 & 43.11 & 59.42 & 11.41 & 35.96 & 41.77 & 60.09 & 11.82 & 36.26 & 42.43 & \textbf{64.02} & \textbf{20.38} & \textbf{42.06} & \textbf{44.76} \\
CodeT5+ & 55.81 & 5.33 & 32.99 & 38.71 & 54.62 & 5.10 & 32.31 & 38.34 & 54.84 & 5.01 & 32.24 & 38.92 & \textbf{56.97} & \textbf{6.25} & \textbf{33.74} & \textbf{39.96} \\

\hline

\multicolumn{17}{c}{Java Corpus} \\
\hline
\multirow{2}{*}{Model} & \multicolumn{4}{c|}{$LM$} & \multicolumn{4}{c|}{$LM$ + $\mathcal{L}_{sft}$} & \multicolumn{4}{c|}{$LM$ + $\mathcal{L}_{drl}$} & \multicolumn{4}{c}{$LM$ + $\mathcal{L}_{sft}$ + $\mathcal{L}_{drl}$} \\
\cline{2-17}
\cline{2-17}
 & Edit-Sim & EM & BLEU-4 & CodeBLEU & Edit-Sim & EM & BLEU-4 & CodeBLEU & Edit-Sim & EM & BLEU-4 & CodeBLEU & Edit-Sim & EM & BLEU-4 & CodeBLEU\\
\hline
GPT-2 [124M] & 57.42 & 4.95 & 32.56 & 42.25 & 56.33 & 4.49 & 31.90 & 42.03 & 52.68 & 4.63 & 32.55 & 42.55 & \textbf{58.58} & \textbf{5.40} & \textbf{33.12} & \textbf{43.68} \\
GPT-2 [1.5B] & 58.14 & 6.24 & 34.24 & 43.47 & 57.27 & 5.86 & 34.00 & 42.71 & 56.89 & 5.47 & 33.88 & 42.35 & \textbf{59.51} & \textbf{6.69} & \textbf{35.02} & \textbf{43.98} \\
CodeGPT & 58.98 & 12.06 & 36.32 & 43.92 & 57.55 & 11.35 & 35.25 & 43.23 & 55.98 & 10.56 & 34.71 & 42.97 & \textbf{59.94} & \textbf{12.57} & \textbf{36.89} & \textbf{44.19} \\
GodeGPT-adapt & 58.87 & 5.57 & 33.53 & 42.97 & 57.24 & 5.17 & 32.88 & 42.66 & 55.38 & 4.80 & 30.88 & 41.78 & \textbf{59.15} & \textbf{5.84} & \textbf{33.68} & \textbf{44.02} \\
CodeGen & 59.33 & 11.52 & 35.96 & 43.08 & 58.66 & 11.03 & 35.24 & 42.71 & 59.14 & 11.41 & 35.66 & 43.29 & \textbf{60.31} & \textbf{12.40} & \textbf{36.99} & \textbf{44.06} \\
StarCoder & 59.94 & 11.88 & 36.24 & 43.56 & 59.18 & 11.03 & 35.75 & 43.22 & 58.68 & 10.71 & 35.18 & 42.91 & \textbf{61.63} & \textbf{13.15} & \textbf{37.96} & \textbf{44.38} \\
CodeT5+ & 53.42 & 4.61 & 31.49 & 36.85 & 53.17 & 4.46 & 31.05 & 35.73 & 52.76 & 4.09 & 30.88 & 35.02 & \textbf{54.11} & \textbf{4.98} & \textbf{32.06} & \textbf{38.61} \\

\hline

\end{tabular}
}
\vspace{-1em}
\end{table}  

\subsection{ Impact of Different Model Learning Objectives (RQ3)} 
In code completion,  SFT primarily aims to maximize the log-likelihood of the next correct code. By contrast, DRL seeks to maximize the reward signal by utilizing a policy-based approach. Due to the different learning goals, we conduct experiments using various combinations of $\mathcal{L}_{sft}$ and $\mathcal{L}_{drl}$ for the model, and the experimental results are shown in Table \ref{table:table3}. Notably, the pre-trained LM in the table has been fine-tuned for 5 epochs using SFT (as indicated in the `$LM$' column).

As shown in Table \ref{table:table3}, when the model is further fine-tuned using only $\mathcal{L}_{sft}$, the loss of the model is further reduced during training. However, the performance of the model on the test set decreases, which is due to the overfitting of the model during training. Secondly, when conducting experiments solely with $\mathcal{L}_{drl}$, we encounter the issue of vanishing gradients during fine-tuning. This phenomenon aligns with the observations in \cite{ranzato2016sequence,bahdanau2016actor,le2022coderl}.
As a result, the performance of the model eventually decreases. However, the performance of the model is further improved when the model is fine-tuned using a combination of $\mathcal{L}_{sft}$ and $\mathcal{L}_{drl}$. First and foremost, SFT concentrates on deciphering the inherent patterns and structures within data, predominantly utilizing vast quantities of labeled datasets. In contrast, DRL is designed to adapt through interactions with its environment, with the goal of optimizing a set reward metric. By merging these two approaches, the model is equipped to navigate dynamic contexts while also capturing the inherent patterns within the data. Hence, when relying exclusively on SFT, the model typically learns based on the loss sourced from labeled data. However, when integrated with DRL, the model draws insights from a more comprehensive feedback system, encompassing aspects such as reward signals, which could lead to improved performance.

\begin{tcolorbox}  
\textbf{Answer to RQ3:} The experimental results show that SFT has a positive facilitating effect on the training of DRL, and the hybrid training strategy of SFT and DRL can alleviate the gradient vanishing problem encountered during the training of DRL, thus effectively improving the performance of the pre-trained LM.
\end{tcolorbox}

\subsection{Quantitative Analysis (RQ4)} 
\begin{figure*}[t]
	\centering
	\begin{minipage}{0.48\textwidth}
		\includegraphics[width=\textwidth]{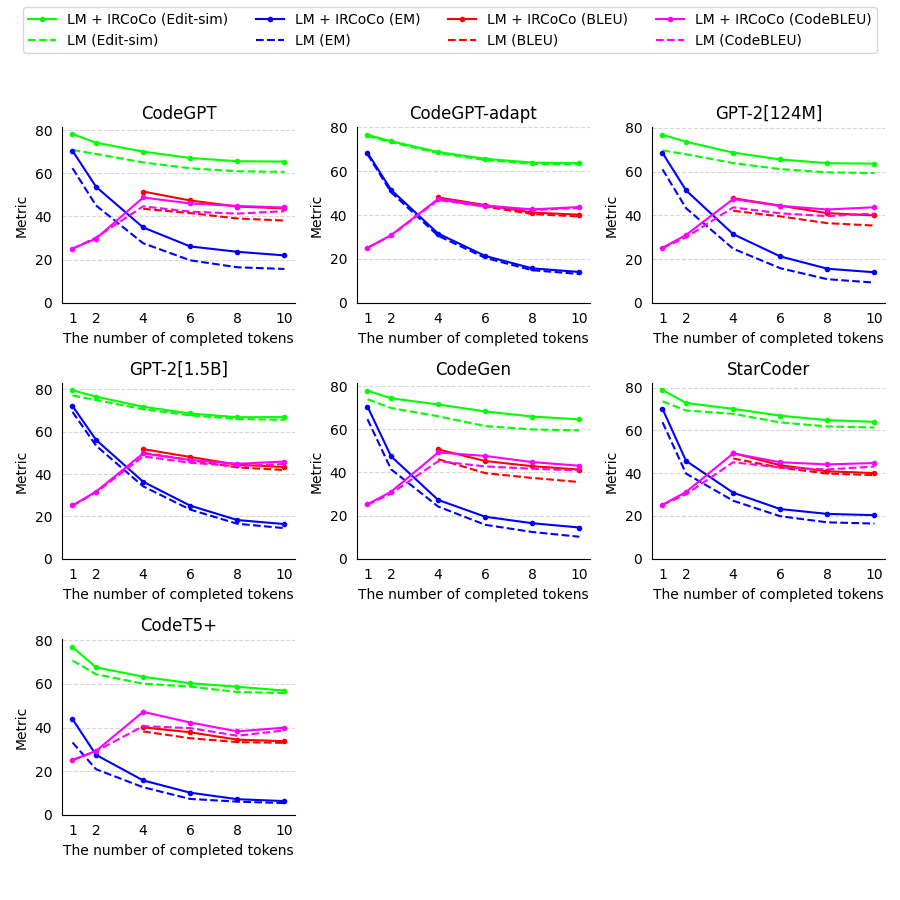}
		\caption{Comparison of the IRCoCo framework under different numbers of tokens (Py150 dataset).}
		\label{line_python}
	\end{minipage}
	\hfill 
	\begin{minipage}{0.48\textwidth}
		\includegraphics[width=\textwidth]{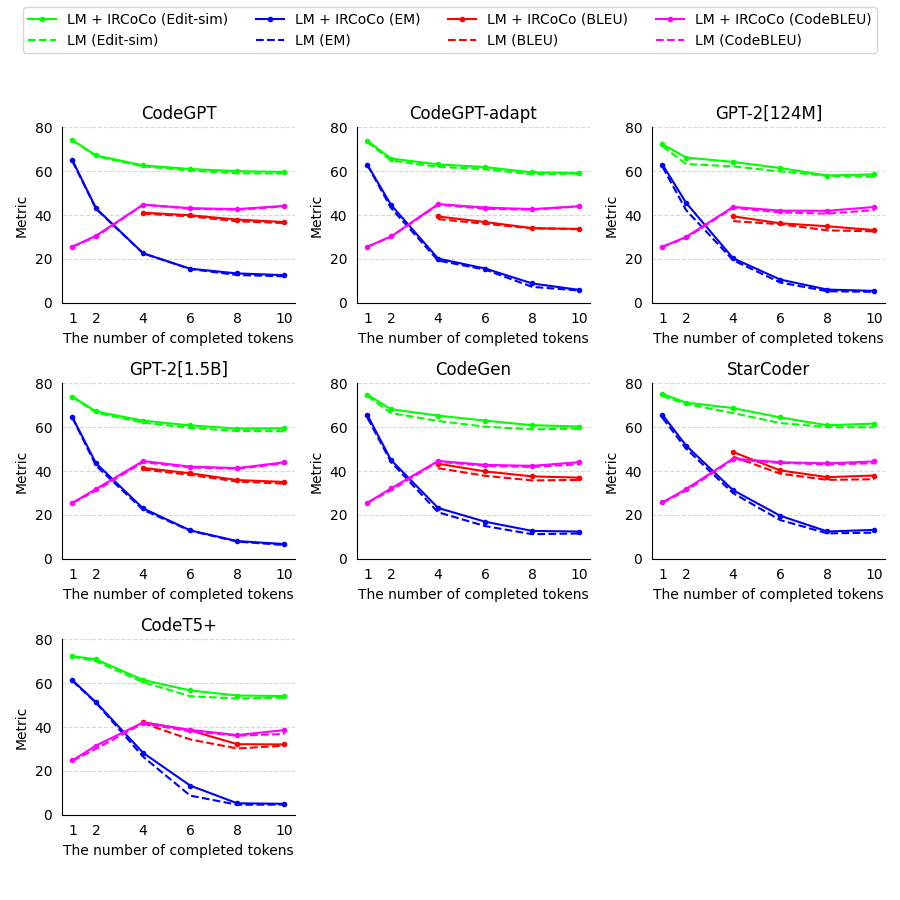}
		\caption{Comparison of the IRCoCo framework under different numbers of tokens (Java Corpus dataset).}
		\label{line_java}
	\end{minipage}
     \vspace{-1em}
\end{figure*}

We analyze the model's completion performance for different numbers of tokens on the Py150 and Java Corpus datasets. Given that we employ BLEU-4, we do not report the BLEU metric for token counts less than 4 in our experiments. The results of these experiments are illustrated in Figures \ref{line_python} and \ref{line_java}. These figures clearly demonstrate that IRCoCo outperforms the pre-trained LMs across different code completion lengths for both the Py150 and the Java Corpus. Particularly notable is IRCoCo's consistent superiority over the pre-trained LMs in cases involving longer completion lengths. This finding suggests that the immediate rewards mechanism enables the model to effectively incorporate the already completed information into its predictions. Furthermore, compared to Python, the improvement brought by IRCoCo when applied to Java has decreased. This limited enhancement can be attributed to Java's exhaustive type system and its intricate syntax, which gives rise to complex coding patterns. Conversely, Python tends to be more streamlined. Given this degree of complexity, DRL necessitates a larger volume of data for effective adaptation and learning. 
Unfortunately, the Py150 dataset is about seven times the size of the Java Corpus dataset, and this discrepancy suggests that there are fewer samples of reinforcement learning data for training in the Java language compared to the Python language, leading to IRCoCo's inferior performance on the Java Corpus compared to its performance on Py150.

\begin{tcolorbox}  
\textbf{Answer to RQ4:} The experimental results show that the pre-trained LM after integrating the IRCoCo framework outperforms the pre-trained LM for different code complement lengths, suggesting that the immediate reward mechanism enables the model to efficiently incorporate already completed information into its predictions.
\end{tcolorbox}

\subsection{Qualitative Analysis (RQ5)}

\begin{figure}[!t]
    \centering
    
    \begin{subfigure}[b]{0.49\linewidth}
        \centering
        \includegraphics[width=\linewidth]{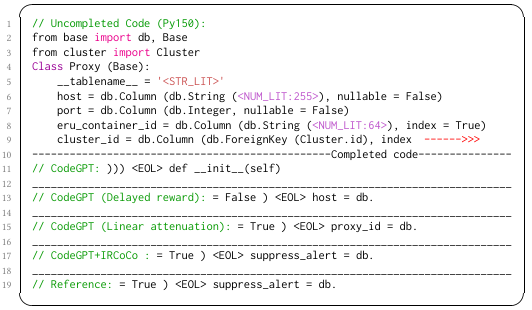}
        \caption{Correct completions (Python)}
        \label{fig:left_top}
        \vspace{-1em}
    \end{subfigure}
    \hfill
    \begin{subfigure}[b]{0.49\linewidth}
        \centering
        \includegraphics[width=\linewidth]{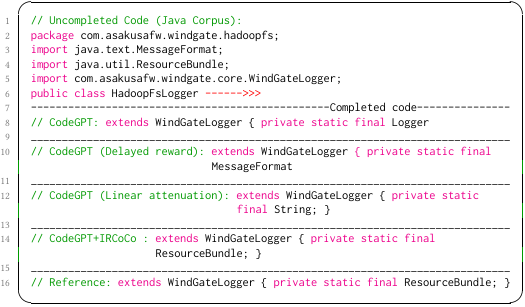}
        \caption{Correct completions (Java)}
        \label{fig:right_top}
        \vspace{-1em}
    \end{subfigure}

    \vspace{0.6cm}
    
    \begin{subfigure}[b]{0.49\linewidth}
        \centering
        \includegraphics[width=\linewidth]{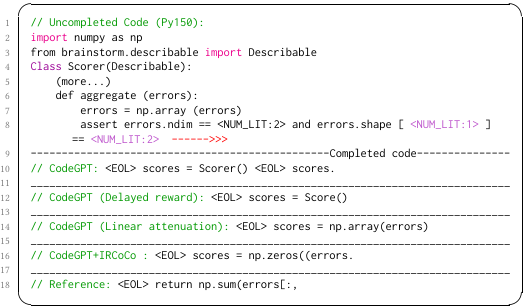}
        \caption{Error completions (Python)}
        \label{fig:left_bottom}
    \end{subfigure}
    \hfill
    \begin{subfigure}[b]{0.49\linewidth}
        \centering
        \includegraphics[width=\linewidth]{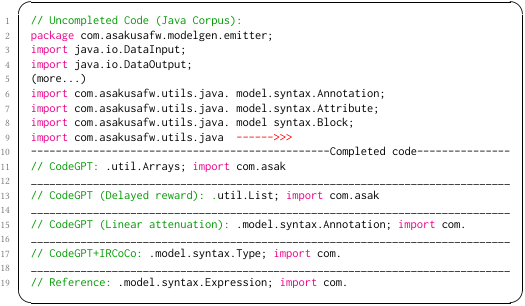}
        \caption{Error completions (Java)}
        \label{fig:right_bottom}
    \end{subfigure}
    \caption{Examples of CodeGPT completion on Py150 and Java Corpus datasets.}
    \label{fig:shili}
    \vspace{-1em}
\end{figure}

Although statistical metrics offer valuable insights, they might not fully reflect the model's predictive capabilities. Thus, we conduct a qualitative assessment of the code generated by IRCoCo. For this analysis, we adopt CodeGPT as the base model and supplement it with diverse techniques to provide a more detailed comparison. Figures \ref{fig:left_top} and \ref{fig:right_top} display cases of successful code completions, while Figures \ref{fig:left_bottom} and \ref{fig:right_bottom} highlight instances where the model failed to generate the desired code.

From our observations, it is evident that when augmented with IRCoCo, CodeGPT exhibits enhanced accuracy in completing incomplete code fragments. In the context of the Py150 dataset, the predictions of CodeGPT were notably inaccurate. When introduced to delayed rewards, the model generated code snippets that, although closer to the reference, were not exact matches. In experiments that utilize LA rewards, we find that the model began to complete \texttt{ = True ) <EOL>}, which shows that the model has developed towards the correct method, but the subsequent variable name \texttt{proxy\_id} is wrong, indicating that rule-based immediate rewards are still not effective enough. Interestingly, the inclusion of IRCoCo's immediate rewards leads to perfect code completion, suggesting the significant role our novel immediate rewards play in guiding the model toward the correct strategy. When using the Java Corpus dataset, we find that the baseline CodeGPT does not perform quite well, its suggestions ceased after the $Logger$ prompt. The completions with delayed and LA rewards both fall short of the reference by one keyword. In contrast, when using the immediate rewards from IRCoCo, the model completes the code correctly. This indicates that our immediate reward mechanism allows the model to adjust its strategy during exploration based on the anticipated benefits of future completions, thus enhancing the accuracy of code completion. In contrast, other immediate reward mechanisms, such as the LA reward scheme with a fixed decay coefficient, cannot accurately perceive the expected returns of future correct completions. Moreover, the coarse-grained (0-1) reward scheme struggles to capture the fine-grained expected returns in code completion tasks. Contrastingly, as evident from Figures~\ref{fig:left_bottom} and \ref{fig:right_bottom}, even when our proposed methodology is not entirely accurate, the results are close enough to the reference that developers may find that they can be used with only minor modifications.

These observations underline the effectiveness of incorporating IRCoCo's immediate rewards into DRL, steering the model to discern the optimal strategy. 
This is largely because our immediate rewards guide the model away from incorrect intermediate solutions and towards best practices during training. 
Additionally, the combined power of SFT and DRL-based alignment, when trained jointly, leverages the strengths of both paradigms, leading to high-quality code.
\begin{tcolorbox}  
\textbf{Answer to RQ5:} The results show that IRCoCo's overall predictions of the model are in the right direction in both datasets, although complementary errors still occur, suggesting that the immediate rewards gradually steer the model away from the wrong solution and towards the right strategy.
\end{tcolorbox}

\section{DISCUSSION}
\noindentparagraph{\textbf{\textup{Why use the current selection of these metrics and why not pass$@k$?}}}
In code generation, generating complete and highly accurate code poses significant challenges. Nevertheless, code completion offers a viable approach that does not require the model to achieve a \SI{100}{\percent} completion rate for correct code. In most cases, the objective is to generate code fragments that are as accurate or similar as possible, allowing programmers to utilize them with minor modifications. To evaluate the code completion model, we have selected Edit-Sim and BLEU as similarity metrics, along with the accuracy metric to demonstrate the model's performance. As the current code completion datasets lack unit test cases, we have not opted to employ pass$@k$~\cite{chen2021evaluating} as the evaluation metric.

\noindentparagraph{\textbf{\textup{Why some relevant models (e.g., CodeRL) cannot do the line-level code completion we are concerned with?}}}
We opted not to draw comparisons with CodeRL~\cite{le2022coderl} due to several factors. Initially, code completion differs fundamentally from code generation. Our study emphasizes line-level code completion, while CodeRL is geared towards generating code from NL descriptions to produce code sequences. Next, the reward mechanism employed by CodeRL does not readily apply to code completion. CodeRL deploys a critic model using an encoder-decoder architecture that provides dense reward signals to the rendered complete code. Under this framework, the encoder receives an NL description, and the decoder outputs complete code that fulfills the functional requirements. In the realm of code generation, the NL description remains unchanged. Conversely, in code completion, the ``context'', which is the code awaiting completion, evolves with each completed token. Furthermore, CodeRL's rewards are based on unit tests. However, since the code completion task does not usually result in creating complete functional code, there are no unit test cases in the datasets. When inspecting code generation datasets endowed with unit tests, such as APPS \cite{hendrycks2021measuring}, it is evident that the code therein addresses programming contests and is not tailored for code completion objectives.

\section{Threats to Validity}

We have identified the following two threats to the validity.

\noindentparagraph{\textbf{\textup{Limitations of the Evaluation Metrics.}}}
Through our experiments, we observe that certain code completion results are presented in the form of ``\texttt{b==a}'' while the actual value is in the ``\texttt{a==b}'' format. Although these two expressions are semantically equivalent in terms of correctness, the way evaluation metrics are computed classifies the completion result as incorrect. In this study, we employ three automated evaluation metrics to assess code completion quality. Despite these metrics inevitably facing the aforementioned issue, we opt for popular and widely adopted metrics to align our evaluations closely with current research directions. Thus, for future work, besides using standard metrics, integrating human evaluations is essential to alleviate this threat.

\noindentparagraph{\textbf{\textup{Metrics for Code Completion Quality Evaluator.}}}
Since manually constructing datasets for training code completion quality evaluators is impractical, we utilized automated evaluation metrics, namely BLEU and Edit-sim, to build our datasets for experiments. However, such exact match-based metrics overlook the semantics of code, which might undermine their reliability in measuring code completion quality. The research community has previously probed the limitations of these metrics~\cite{evtikhiev2023out,roy2021reassessing}, yet they remain a formidable challenge. For quality evaluators in code completion, the acceptance rate of completions stands as the most desirable metric~\cite{sun2023don}. Obtaining such data, however, is challenging. Given that in the industrial domain, code completions are often short and do not need to be entirely correct—just requiring minor modifications by the programmer—we opted for similarity metrics that align closely with the ideal acceptance rate. To address these concerns, we call upon a broader participation of researchers to further delve into and advance studies in this realm.

\section{Related Work}

\noindentparagraph{\textbf{\textup{Code Completion.}}}
The code completion task is a critical task in the field of software engineering, and early code completion methods typically employ rule-based heuristic approaches and statistical-based techniques. The heuristic rule-based approach~\cite{stubenschrott2005context,bruch2009learning} for code completion involves manually writing rules to generate code suggestions. However, this approach requires a significant amount of manual labor and expertise and may struggle to cope with complex code completion tasks. The statistical-based approach~\cite{allamanis2013mining,bettenburg2015towards} for code completion employs N-gram LMs to model source code and learn code completion models by statistically analyzing large amounts of code data. However, this approach requires high-quality and high-quantity data and may struggle to handle semantic and contextual information in the code.

With the continuous development of deep learning techniques, SFT-based LMs have made significant strides in code completion. Liu et al.~\cite{liu2016neural} proposed a neural network-based approach to generate more accurate completion suggestions by learning contextual information about the code. Li et al.~\cite{li2022cctest} proposed CCTEST, a code completion framework for testing and repairing, which repairs the code after completion as the final output. Li et al.~\cite{li2018code} proposed a pointer hybrid network to solve the out of vocabulary (OOV) problem, which improved the generalization ability and robustness of the model. In addition to the above-mentioned approach of treating source code sequences as code token sequences, some researchers have proposed transforming source code into an abstract syntax tree to predict the next node of the tree. This approach can better capture the semantic and structural information in the code, thus improving the effectiveness and quality of code completion. With the recent rise of pre-trained LMs, several research efforts have begun to utilize these models to solve the code completion problem~\cite{lu2021codexglue}. This approach leverages the linguistic representation capabilities learned from large amounts of data to achieve the code completion task by fine-tuning on the datasets. Although these methods have good performance at first, \textit{exposure bias} still needs to be addressed in SFT.

\noindentparagraph{\textbf{\textup{Sequence Generation via Reinforcement Learning.}}}
Related to code completion is the field of sequence generation, where DRL methods are often used to address the impact of \textit{exposure bias} present in SFT. In a previous study on DRL,  Ranzato et al.~\cite{ranzato2016sequence} directly used the final optimized indicators, BLEU and ROUGE, as reward signals and then used DRL algorithm to optimize network parameters in machine translation tasks. Wan et al.~\cite{wan2018improving,wang2020reinforcement} trained a generative model in the code summarization task using the actor-critic algorithm and a reward function consisting of BLEU metrics. Reed et al.~\cite{reed2016learning} used the DRL algorithm to train policy networks in image generation tasks by simulating the drawing process using the policy gradient method. Recently, Chen et al.~\cite{chen2021decision} proposed Decision Transformer, a method that transforms the DRL problem into a conditional sequence modeling problem by converting state and action sequences into inputs for a conditioned LM, which is then trained using the Transformer architecture to achieve the DRL task. In the field of code completion, Shojaee et al.~\cite{shojaee2023execution} proposed a novel reward function that combines discrete compiler feedback with the syntactic and semantic match scores between the generated code and the executable target. This approach aims to reduce the sparsity of the reward function by combining multiple metrics to better guide the generation of code that more closely aligns with the correct target.

Traditional reward designs in DRL have focused on optimizing evaluation metrics directly after sequence generation or relying on the functional correctness of unit tests for assignment. However, such delayed reward mechanisms often result in slow model convergence and a higher likelihood of getting trapped in a local optimal solution. In the context of code completion, unit tests cannot be used as a direct reward signal since the generated code may not necessarily form a complete code fragment. Therefore, it is crucial to devise a mechanism that provides immediate rewards for the sequence generation process.

\noindentparagraph{\textbf{\textup{Large Language Models (LLMs).}}}
At present, several closed-source models, including ChatGPT~\cite{r7OpenAI2022} and GPT-4~\cite{OpenAI_2023}, as well as open-source models such as CodeGen~\cite{nijkamp2022codegen}, Code Llama~\cite{roziere2023code}, CodeT5+~\cite{wang2023codet5plus}, and StarCoder~\cite{li2023starcoder}, have showcased impressive capabilities in NL2Code tasks. 
Our primary objective is not to surpass these LLMs. Instead, we introduce a fine-tuning mechanism tailored for LM-based code completion. Theoretically, this mechanism can be applied across models constrained by exposure bias and delayed reward challenges.

\section{Conclusion}

We present IRCoCo, a DRL framework tailored for code completion, which collaboratively employs SFT fine-tuning and DRL-based alignment to augment pre-trained LMs. Specifically, we introduce a code completion quality evaluator to provide immediate rewards for the code completion generation process. We evaluate IRCoCo in combination with pre-trained LMs, and our comprehensive analysis demonstrates that IRCoCo can improve the performance of pre-trained LMs. In essence, IRCoCo is an innovative framework that leverages immediate rewards to improve pre-trained LMs' performance on the code completion task through DRL.

\noindentparagraph{\textbf{\textup{Data Availability.}}}
All experimental data and source code used in this paper are available at \texttt{\url{https://github.com/Libolun-star/IRCoCo}}.

\section*{Acknowledgments}
The work is supported in part by the Natural Science Foundation of Shandong Province, China (Grant No. ZR2021MF059), the National Natural Science Foundation of China (Grant Nos. 62192731, 62072007, 62192733, 61832009, 62192730), the National Key R\&D Program under (Grant No. 2023YFB4\newline503801) and the Key Program of Hubei (Grant No. JD2023008).

\bibliographystyle{ACM-Reference-Format}



\end{document}